\documentclass[12pt,showpacs]{revtex4}
\usepackage[english]{babel}
\usepackage{graphicx}
\usepackage{epsfig}
\def\be{\begin{equation}}
\def\ee{\end{equation}}
\def\bea{\begin{eqnarray}}
\def\eea{\end{eqnarray}}
\def\nn{\nonumber}

\def\ket#1{\hbox{$\vert #1\rangle$}}   
\def\bra#1{\hbox{$\langle #1\vert$}}   
\def\oneh{{\textstyle {1\over 2}}}

\def          
\simeq{{\ \lower2pt\hbox{$-$}\mkern-13mu \raise2pt \hbox{$\sim$}\ }} 

\def\bold#1{\setbox0=\hbox{$#1$}%
      \kern-.02em\copy0\kern-\wd0
      \kern.04em\copy0\kern-\wd0
      \kern-.02em\raise.0433em\box0 }
\def\smbold#1{\setbox0=\hbox{$\scriptstyle#1$}%
      \kern-.02em\copy0\kern-\wd0
      \kern.04em\copy0\kern-\wd0
      \kern-.02em\raise.0433em\box0 }

\newcommand{\lcvec}[3]{\left[\;#1\;,\;#2\;,\;#3\;\right]}
\def\oggi{\number\day.\space 
\ifcase\month\or
 1.\or 2.\or 3.\or 4.\or 5.\or 6.\or
 7.\or 8.\or 9.\or 10.\or 11.\or 12.\fi
 \space\number\year}

\begin{document}
%
%
%


\title{Electroweak structure of the nucleon, meson cloud and light-cone 
wavefunctions}

\author{B.~Pasquini and S.~Boffi
}
\affiliation{Dipartimento di Fisica Nucleare e Teorica, Universit\`a degli
Studi di Pavia and INFN, Sezione di Pavia, Pavia, Italy}
\date{\today}
\begin{abstract}
{The meson-cloud model of the nucleon consisting of a system of three valence quarks surrounded by a meson cloud is applied to study the electroweak structure of the proton and neutron. Light-cone wavefunctions are derived for the dressed nucleon as pictured to be part of the time a bare nucleon and part of the time a baryon-meson system. Configurations are considered where the baryon can be a nucleon or a $\Delta$ and the meson can be a pion as well as a vector meson such as the $\rho$ or the $\omega$. An overall good description of the electroweak form factors is obtained. The contribution of the meson cloud is small and only significant at low $Q^2$. Mixed-symmetry $S'$-wave components in the wavefunction are most important to reproduce the neutron electric form factor. Charge and magnetization densities are deduced as a function of both the radial distance from the nucleon center and the transverse distance with respect to the direction of the three-momentum transfer. In the latter case a central negative charge is found for the neutron. The up and down quark distributions associated with the Fourier transform of the axial form factor have opposite sign, with the consequence that the probability to find an up (down) quark with positive helicity is maximal when it is (anti)aligned with the proton helicity.
}
\end{abstract}
\maketitle 
%


\section{Introduction}
\label{sect:intro}

Since the discovery of the proton finite size by Hofstadter and coworkers more 
than 50 years ago~\cite{Hofstadter}, electromagnetic form factors have played 
a privileged role in the investigation of the nucleon structure (for recent 
reviews, see~\cite{Gao03,Hyde04,ADZ06,PPV06,Kees06}). 

In principle, these quantities reflect the strong  interaction between quarks 
inside the nucleon and should be described by quantum chromodynamics (QCD). 
In recent years important progress has been made with lattice QCD 
simulations~\cite{QCDSF05b,QCDSF06b,AKNT06,Adelaide06,LHPC06a}. Generic features of the baryon octet mass spectrum are reproduced well in quenched lattice QCD simulations (in which sea-quark contributions are neglected) and electromagnetic properties can be studied with pion masses as low as 0.3 GeV~\cite{Adelaide06}.  Qualitative agreement with the experimental data has been obtained~\cite{QCDSF05b,QCDSF06b}, e.g. the flavour dependence of the Dirac form factor $F_1(Q^2)$~\cite{Jakob05}. The isovector nucleon form factors were calculated in both the quenched and unquenched approximations with configurations for pion masses down to 
380 MeV~\cite{AKNT06}. Small  unquenching effects and results larger than the 
experimental data were found. Evaluating the axial form factors as well as the $\pi NN$ and $\pi N\Delta$ form factors  it was possible to check the Goldberger-Treiman relations~\cite{AKNT07}. Qualitatively consistent results with experiment were obtained for the isovector form factor ratio $G_P(Q^2)/G_A(Q^2)$ of the nucleon axial vector form factors~\cite{LHPC06a}. At present, the state of the 
art is still limited by systematic errors which are related to the fact that 
calculations are performed on finite volumes, at finite lattice spacings and 
at quark masses which are still relatively large. In addition, the 
extrapolation to the chiral limit with the help of the chiral effective field 
theory requires calculations of higher order than presently 
available~\cite{Bernard98,Leinweber01,Hemmert02} (see also~\cite{MSed06}) in 
order to account for the non-analytic behaviour of the form factors on the 
quark masses~\cite{Ashley04}.

Therefore, model calculations are still a valuable tool for understanding the
 internal dynamics of the nucleon. In particular, many theoretical calculations
 have been done to investigate possible interpretations of the decreasing 
ratio of the proton electric to magnetic form factors, 
$\mu_p G_E^p(Q^2)/G_M^p(Q^2)$, as a function of $Q^2$ 
(see, e.g., \cite{Punjabi05} for an overview and ~\cite{Carlson07,Arrington07} for the two-photon physics in  elastic electron scattering) and we refer the reader to 
Refs.~\cite{ADZ06,PPV06} for a discussion of the different interpretations of nucleon electromagnetic form factors proposed in the literature.

Here we want to consider the possibility that the physical nucleon is a bare nucleon surrounded by a meson cloud as a consequence of the 
spontaneously-broken chiral symmetry. As first discussed in the context of 
deep inelastic scattering~\cite{FeyPHI,Sull72}, a pion cloud can give an 
explanation of the flavor-symmetry violation in the sea-quark distributions of
 the nucleon thus accounting for the violation of the Gottfried sum 
rule~\cite{Thomas83}. This cloud will manifest itself as an extension of the
 charge distribution of protons and neutrons, which should be observable in 
the electromagnetic form factors at relatively small values of $Q^2$. In fact,
  the neutron charge density extracted from available data shows a positive 
core surrounded by a negative surface charge, peaking at just below 1 fm, 
which can be attributed to a negative pion cloud~\cite{Kelly02}. This is 
confirmed by the analysis of Ref.~\cite{FW03} showing a pronounced bump 
structure in the neutron electric charge form factor $G_E^n(Q^2)$ (and a dip in  the other nucleon form factors) around $Q^2=0.2-0.3$ GeV$^2$, which can
 be interpreted as a signature of a very long-range contribution of the pion
 cloud extending out to 2 fm. It must be said, however, that from dispersion
 relation analysis~\cite{Hammer04a,Hammer04b} the pion cloud should peak much 
more inside the nucleon, at $\sim 0.3$ fm, and the desired bump-dip structure 
of Ref.~\cite{FW03} can only be achieved at the cost of low-mass poles close 
to the $\omega$ mass in the isoscalar channel and to the three-pion threshold 
in the isovector channel~\cite{BHM07}. In addition, while confirming the 
long-range positively (negatively) charged component of the proton (neutron)
 charge density, a recent model-independent analysis of the infinite-momentum-frame charge density of partons in the transverse plane~\cite{Miller07} is suggesting that the neutron parton charge density is negative at the center.
 
Mesonic degrees of freedom are naturally taken into account in the baryon chiral perturbation theory that is the effective field theory of the standard model at low energies and small momentum transfer. The electromagnetic form factors of the nucleon have been calculated to  fourth order (one-loop) in baryon chiral perturbation theory within the manifestly Lorentz-invariant infrared regularization approach~\cite{Kubis} and the extended on-mass-shell renormalization scheme~\cite{Fuchs04,Schindler05}. The inclusion of vector mesons as explicit degrees of freedom results in a considerably improved description, accurate up to $Q^2\simeq 0.4$ GeV$^2$.

The problem of considering the meson cloud surrounding a system of three
 valence quarks has been addressed already in the past in a variety of 
papers (see, e.g., 
\cite{Schutte,DHSS97,LThW98,DFS99,FGLP06,Julia06,Riska07,Giannini07,Melo07} 
and references therein). Along the lines originally proposed in Refs.~\cite{drelllevyyan, Zoller92a, Zoller92b} and developed in~\cite{DHSS97}, in this paper a baryon-meson Fock-state expansion is used to construct the state $\ket{\tilde N}$ of the physical nucleon. In the one-meson approximation the state $\ket{\tilde N}$ is pictured as being part of the time a bare nucleon, $\ket{N}$, and part of the time a baryon-meson system, $\ket{BM}$. The bare nucleon is formed by three valence quarks identified as constituent quarks according to the ideas discussed 
in~\cite{JaRo}. The model was revisited in Ref.~\cite{PB06} to study generalized parton distributions where the meson cloud gives an essential contribution in the socalled ERBL region. We apply here the model to calculate the electroweak form factors of the nucleon. The baryon-meson system is assumed to include configurations where the baryon can be a nucleon or a Delta and the meson can be a pion as well as a vector meson such as the $\rho$ or the $\omega$.

In Sect.~\ref{wf_mcm} the relevant formulae of the meson-cloud model are collected and the light-cone wavefunctions derived. The calculation of the electroweak form factors of the proton and neutron is illustrated in Sect.~\ref{sect:ff}, and the results are presented and discussed in Sect.~\ref{sect:results}. Concluding remarks are collected in the final section. Technical details necessary to calculate the vertex functions describing the transition to a baryon-meson state with vector mesons are given in Appendix~\ref{appendixa}.


\section{Light-cone Wavefunction of the nucleon in the meson cloud model}
\label{wf_mcm}

The derivation of the nucleon light-cone wavefunction (LCWF) in the meson cloud model
has been already discussed in Ref.~\cite{PB06}.
In this section we review some pertinent formulae necessary for the calculation
of the form factors.
In the meson cloud model the nucleon is viewed as a quark core, 
termed the bare nucleon, surrounded by a meson cloud.
The mesonic effects are treated perturbatively, by truncating the Fock-space
expansion of the nucleon state to the dominant components given by
the bare nucleon and the state containing a virtual meson  
with a recoiling baryon.
The corresponding quantum state of the physical nucleon ($\tilde N$), with four-momentum 
$p_N^\mu=(p^-_N,p^+_N,{\mathbf p}_{N\perp})\equiv(p^-_N,\tilde p_N)$ and helicity $\lambda$, can be written as 
\begin{eqnarray}
|\tilde p_N,\lambda;\tilde N\rangle
& &
=\sqrt{Z}|\tilde p_N,\lambda; N\rangle
+
\sum_{B,M}
\int \frac{{\rm d}y{\rm d}^2{\mathbf k}_{\perp}}{2(2\pi)^3}\,
\frac{1}{\sqrt{y(1-y)}}
\sum_{\lambda',\lambda''}
\phi_{\lambda'\lambda''}^{\lambda \,(N,BM)}(y,{\mathbf k}_\perp)\nonumber\\
& &\quad{}\times
|yp^+_N,{\mathbf k}_{\perp}+y{\mathbf p}_{N\perp},\lambda';B\rangle\,
|(1-y)p^+_N,-{\mathbf k}_{\perp}+(1-y){\mathbf p}_{N\perp},\lambda'';M\rangle,
\label{eq:14}
\end{eqnarray}
where the function 
$\phi_{\lambda'\lambda''}^{\lambda\,(N,BM)}(y,{\mathbf k}_\perp)$ is the probability amplitude to find a physical nucleon with helicity $\lambda$
in a state consisting of a  virtual baryon $B$ and a virtual meson $M$,
 with the baryon having helicity $\lambda'$, longitudinal 
momentum fraction $y$ and transverse momentum ${\mathbf k}_\perp$, and the 
meson having helicity $\lambda''$, 
longitudinal momentum fraction $1-y$ and 
transverse momentum $
-{\mathbf k}_\perp$.
From the normalization condition of the nucleon state
\be
\langle \tilde p'_N,\lambda';\tilde N
|\tilde p_N,\lambda;\tilde N\rangle
=2(2\pi)^3 p^+_N\delta(p'^+_N-p^+_N)\delta^{(2)}({\mathbf p}'_{N\perp}-{\mathbf p}_{N\perp})
\delta_{\lambda\lambda'},
\ee
one obtains the following condition on the renormalization factor $Z$
\be
1=Z+\sum_{B,M} P_{BM/N},
\label{eq:norm}
\ee
with
\be
P_{BM/N}=
\int \frac{{\rm d}y{\rm d}^2{\mathbf k}_{\perp}}{2(2\pi)^3}\,
\sum_{\lambda',\lambda''}
|\phi^{1/2(N,BM)}_{\lambda'\lambda''}(y,{\mathbf k}_\perp)|^2.
\label{eq:prob}
\ee
From the definitions in Eqs.~(\ref{eq:norm}) and (\ref{eq:prob}), one can interpret the factor $Z$ as the probability of finding a bare nucleon in the physical nucleon, and $P_{BM/N}$ as the probability of fluctuation of the nucleon in a baryon-meson state.

The probability amplitude $\phi^{\lambda(N,BM)}_{\lambda'\lambda''}$ 
can be calculated using time-ordered perturbation theory in the 
infinite-momentum frame as explained in Ref.~\cite{PB06}. 
The final result reads 
\begin{eqnarray}
\phi_{\lambda'\lambda''}^{\lambda\,(N,BM)}(y,{\mathbf k}_\perp)
=\frac{1}{\sqrt{y(1-y)}}\,
\frac{V^\lambda_{\lambda',\lambda''}(N,BM)}
{M^2_N-M^2_{BM}(y,{\mathbf k}_\perp)},\,
\label{eq:15}
\end{eqnarray}
where $V^\lambda_{\lambda',\lambda''}(N,BM)$ is the vertex function 
describing the transition of the nucleon into a baryon-meson state, with 
squared invariant mass 
\begin{equation}
M^2_{BM}(y,{\mathbf k}_\perp)\equiv\frac{M_B^2+{\mathbf k}^2_\perp}{y}
+\frac{M_M^2+{\mathbf k}^2_\perp}{1-y}.
\label{eq:emmedue}
\end{equation}

The vertex function $V^{\lambda}_{\lambda',\lambda''}(N,BM)$ 
has the following 
general expression~\cite{Speth98}
\be
V^{\lambda}_{\lambda',\lambda''}(N,BM)=
\bar u_{N \alpha}(\tilde p_N,\lambda) v^{\alpha\beta \gamma}
\chi_\beta(\tilde p_M,\lambda'')\psi_\gamma(\tilde p_B,\lambda'),
\label{eq:13}
\ee
where $u_N$ is the nucleon spinor, $\chi$ and $\psi$ are the field operators
of the intermediate meson and baryon, respectively, and
$\alpha,$ $\beta$, $\gamma$ are bi-spinor and/or vector indices depending
on the representation used for particles of given type.
The explicit expressions 
 for the $\pi N$ and $\pi \Delta$ cases have been derived in Appendix C of 
Ref.~\cite{PB06}, while the corresponding results for transitions with vector 
mesons are  
worked out in Appendix~\ref{appendixa}.
Because of the extended structure of the hadrons involved, one has also 
to multiply the coupling constant 
for pointlike particles in the interaction operator
 $v^{\alpha\beta \gamma}$
by phenomenological vertex form factors.
These form factors  parametrize the unknown 
microscopic effects at the vertex and have to obey the constraint
$F_{NBM}(y,k_\perp^2 )=F_{NBM}(1-y,k_\perp^2 )$ to ensure basic properties 
like charge and momentum conservation simultaneously ~\cite{MST99}.
To this aim we will use the following functional form 
\begin{equation} 
F_{NBM}(y,k_\perp^2 )= \mbox{exp}
\left[\frac{M_N^2-M^2_{BM}}{2\Lambda_{BM}}\right],
\label{eq:74}
\end{equation}
where $\Lambda_{BM}$ is a cut-off parameter.

For the hadron states of the bare nucleon and baryon-meson components in 
Eq.~(\ref{eq:14}) we adopt a light-cone constituent quark model,
 by using the minimal Fock-state wavefunction in the light-cone formalism, i.e.
\be
\ket{{\tilde{p}_H},\lambda;H} = \sum_{\tau_i,\lambda_i}
\int\left[\frac{{\rm d}x }{\sqrt{x}}\right]_N
[{\rm d}^2{\mathbf k}_\perp]_N
\Psi_\lambda^{H,[f]}(\{x_i,{\mathbf k}_{\perp i};\lambda_i,\tau_i\}_{i=1,...,N})
\prod_{i=1}^N
\ket{x_i p^+_H, \, {\mathbf p}_{i\perp},\lambda_i,\tau_i},
\label{eq:18}
\ee
where 
$\Psi_\lambda^{H,[f]}(\{x_i,{\mathbf k}_{\perp i};\lambda_i,\tau_i\}_{i=1,..,N})$ is the momentum LCWF  which gives the probability amplitude for finding in the hadron $N$ partons with momenta
 $(x_i p^+_H, {\mathbf p}_{i\perp}=\,{\mathbf k}_{i\perp} + x_i {\mathbf p}_{H\perp})$, 
and spin and isospin variables $\lambda_i$ and $\tau_i,$ respectively. 
In Eq.~(\ref{eq:18}) and in the following formulae, the integration measures are defined by
\be
\label{eq:19}
\left[\frac{{\rm d} x}{\sqrt{x}}\right]_N 
= \left(\prod_{i=1}^N \frac{{\rm d} x_i}{\sqrt{x_i}}\right)
\delta\left(1-\sum_{i=1}^N x_i\right),
\ee
\be
\label{eq:20}
[{\rm d}^2{\mathbf k}_\perp]_N = \left(\prod_{i=1}^N
\frac{{\rm d}^2{\mathbf k}_{\perp\,i}}{2(2\pi)^3}\right)\,2(2\pi)^3\,
\delta\left(\sum_{i=1}^N {\mathbf k}_{\perp\,i}\right),
\ee
where the number of valence partons is $N=3$ and $N=2$ for the baryon and 
meson case, respectively.
As explained in Ref.~\cite{BPT}, the wavefunction $\Psi_\lambda^{H,[f]}$ can be obtained by transforming the ordinary equal-time (instant-form) wavefunction in the rest frame into that in the light-front dynamics, by taking into account relativistic effects such as the Melosh-Wigner rotation, i.e.
\bea
\label{eq:psih}
 \Psi_\lambda^{H,[f]}
(\{x_i,{\mathbf k}_{\perp i};\lambda_i,\tau_i\}_{i=,1,..,N})  
& = &2(2\pi)^3\frac{1}{\sqrt{M_0}}
\prod_{i=1}^N
\left(\frac{\omega_i}
{x_i}\right)^{1/2}\nn\\
& &\times
\sum_{\mu_1,...,\mu_N}
\Psi_\lambda^{H,[c]}(\{{\mathbf k}_i;\mu_i,\tau_i,\mu_i\}_{i=1,...,N})
\prod_{i=1}^N {D}^{1/2\,*}_{\mu_i\lambda_i}(R_{cf}({ \tilde k}_i)),\nonumber\\
& &
\label{eq:75}
\eea 
where $\Psi_\lambda^{H,[c]}$ is the canonical wavefunction,  
and ${D}^{1/2\,*}_{\mu_i\lambda_i}(R_{cf}({\tilde k}_i))$ are the Melosh
 rotations defined in Ref.~\cite{BPT}.
In Eq.~(\ref{eq:75}), $\omega_i=\sqrt{m^2+{\mathbf k}_i^2}$ is the energy
of the $i$-th quark, and $M_0=\sum_i \,\omega_i$ is the free mass
of the system of $N$ non-interacting quarks.

In our model calculation, we take into account the meson cloud 
contribution corresponding to $\pi$, $\rho$, and $\omega$, with the 
accompanying
baryon in the $\ket{BM}$ component  of the dressed nucleon being a nucleon 
or a $\Delta$.
The instant-form wavefunction is  constructed as the 
product of a momentum wavefunction, which is spherically symmetric and 
invariant under permutations, and a spin-isospin wavefunction, which is 
uniquely determined by SU(6)-symmetry requirements.

In the case of the nucleon, we adopt the momentum wavefunction of
Ref.~\cite{Schlumpf94a}, which reads
\bea
\label{eq:psifc}
\psi^{N,[c]}(\{{\mathbf k}_i\}_{i=1,3})
=\frac{N'}{(M_0^2+\beta^2)^\gamma},
\label{eq:76}
\eea 
with $N'$  a normalization factor. In Eq.~(\ref{eq:76}), the scale $\beta$,
the parameter $\gamma$ for the power-law behaviour, and the quark mass $m$ are 
taken as free parameters, and will be determined by a comparison with experimental 
data  as explained in Sect.~\ref{sect:results}.

The $\Delta$ is described as a state of isospin $T=3/2$ obtained as a pure 
spin-flip excitation of the nucleon,  with the corresponding momentum 
wavefunction equal to that of the nucleon in Eq.~(\ref{eq:76}).

Furthermore, the canonical wavefunction of the pion is taken 
from Ref.~\cite{Choi99} and reads 
\begin{eqnarray}
\psi^{\pi,[c]}(\vec{k}_1,\vec{k}_2)
=\frac{i}{\pi^{3/4}\alpha^{3/2}}
\exp{[-k^2/(2\alpha^2)]},
\label{eq:can_psi}
\end{eqnarray}
with $\vec k=\vec k_1=-\vec k_2$,  and the two parameters 
$\alpha=0.3659$ GeV and $m_q=0.22$ GeV 
fitted to the pion form factor data.
The phase of the pion wavefunction (\ref{eq:can_psi}) 
is consistent with that of the antiquark spinors of Ref.~\cite{Brodsky:1989pv}.

The wavefunction of the $\rho$ differs from the pion only in the spin 
component, with the rest-frame spin states of the $q\bar q$ pair coupled to $J=1$ instead of $J=0$.
Similarly, the $\omega$ is described by the same spin and momentum 
wavefunction as the $\rho$, 
but with the isospin component corresponding to a singlet state.
This choice corresponds to assuming an ideal mixing in the vector sector, 
since the effects of the $\phi-\omega$ mixing are irrelevant in the 
calculation of the  meson cloud contribution to the nucleon form factors.
For the same reason, also the effects of the $\rho^0-\omega$ mixing are 
neglected.

Finally, we need to specify the parameters entering in the vertex functions.
The cutoff $\Lambda_{BM}$ in Eq.~(\ref{eq:74}) should in principle  be 
different for each $BM$ component. However, the J\"ulich 
group~\cite{HSS96} and Zoller~\cite{Zoller92a} used high-energy particle 
production data  to determine all the $\Lambda_{BM}$ of interest, and found  
that the data could be described by two parameters: $\Lambda_1$ for octet 
baryons and pseudoscalar and vector mesons, 
and $\Lambda_2$ for decuplet baryons. 
We have chosen the values
$\Lambda_1=0.61$ GeV and $\Lambda_2=0.81$ GeV, which are 
consistent with the ones adopted in the cloudy bag model~\cite{MST99} to  
obtain a good fit to both the violation of the Gottfried sum rule and 
the measured sea quark contribution in the unpolarized parton distribution. 
For the $NBM$ coupling constants at the interaction vertex we used the 
numerical values given in Refs.~\cite{BoTho99,Machleidt} in the case of the 
$\pi$ and the $\rho$. Instead, for the $NN\omega$ coupling, $g_{NN\omega}$,
we used the result from the analysis of Ref.~\cite{Alberg} 
about the $\omega$ contribution to the unpolarized  anti-quark distributions,
which favors a much smaller value for $g_{NN\omega}$
than the one used to describe the nucleon scattering data.
The numerical values for each of the $BM$ states are summarized in Table~\ref{table1}.

\begin{table}
\caption{\small  Coupling constants at the $NBM$ interaction vertex.}
\label{table1} 
\begin{center}
\begin{tabular}{c|c|c|c|c|c}
\hline\hline
$\frac{g^2_{NN\pi}}{4\pi}$ 
& $\frac{f^2_{N\Delta\pi}}{4\pi}$ 
& $\frac{g^2_{NN\rho}}{4\pi^2}$
& $f_{NN\rho}$
&  $\frac{f^2_{N\Delta\rho}}{4\pi}$ 
&$\frac{g^2_{NN\omega}}{4\pi^2}$
\\\hline
$13.6$ 
& $11.08$ GeV$^{-2}$
& $0.84$
&$6.1 \, g_{NN\rho}$
&$20.45$ GeV$^{-2}$
&8.1
\\
\hline\hline
\end{tabular}
\end{center}
\end{table}

Finally, with the specified parameters, the probabilities for each of the $BM$ component
in the dressed nucleon are
\begin{eqnarray*}
& & P_{N \pi /p}=P_{p \pi^0 /p}+P_{n\pi^+ /p}=3P_{p \pi^0 /p}= 5.1\%,\\ 
& & P_{\Delta \pi/p}=P_{\Delta^{++} \pi^-/p}+P_{\Delta^+ \pi^0/p}
+P_{\Delta^0 \pi^+/p}=2P_{\Delta^{++} \pi^-/p}=3.40\%,\\
& & P_{N \rho /p}=P_{p \rho^0 /p}+P_{n\rho^+ /p}=3P_{p \rho^0 /p}= 0.11\%,\\ 
& & P_{\Delta \rho/p}=P_{\Delta^{++} \rho^-/p}+P_{\Delta^+ \rho^0/p}
+P_{\Delta^0 \rho^+/p}=2P_{\Delta^{++} \rho^-/p}=0.67\%,\\
& & P_{N \omega /p}=P_{p \omega/p}= 0.013\%.
\end{eqnarray*}


\section{Electroweak form factors of the nucleon in the meson-cloud model}
\label{sect:ff}

\begin{figure}[ht]
\begin{center}
\epsfig{file=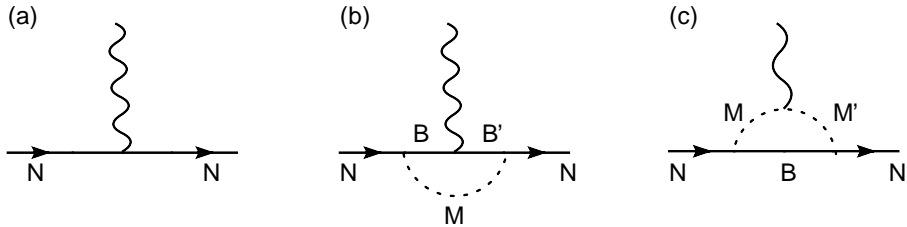,  width=12cm}
\end{center}
\caption{\small Electromagnetic interaction vertex for a bare nucleon (a), and virtual baryon (b) and meson (c) components of a dressed nucleon.}
\label{fig:fig1}
\end{figure}

The Dirac and Pauli form factors $F_1(Q^2)$ and $F_2(Q^2)$ of the nucleon are given by the spin conserving and the spin-flip matrix elements of the vector current $J_V^+=J_V^0+J_V^3$ 
\bea
F_1(Q^2)&=&\langle \tilde p+\tilde q, \frac{1}{2}|J^+_V|\tilde p,\frac{1}{2}
\rangle,\\
(q_x+iq_y)F_2(Q^2)&=&2 M_N 
\langle \tilde p+\tilde q, -\frac{1}{2}|J^+_V|\tilde p,\frac{1}{2}\rangle,
\eea
where $Q^2=-q^2$.
As was first shown by Drell and Yan~\cite{DY}, the calculation of the 
form factors
is conveniently done in a coordinate frame where $q^+=0$. 
In particular, we will use a symmetric frame where the nucleon momenta are
given by 
\begin{eqnarray} 
p_N  &=& 
\lcvec{\frac{M^2_N+{\mathbf q}_\perp^2/4}{p_N^{\,+}}}
      { p_N^{\,+}}
      {-\frac{{\mathbf q}_\perp}{2}} 
      \equiv \left[\frac{M^2_N+{\mathbf q}_\perp^2/4}{p_N^{\,+}}
\,,{\tilde p}_N\right], \nn \\
p_N^{\prime} &=& 
\lcvec{\frac{M^2_N+{\mathbf q}_\perp^2/4}{\bar p_N^{\,+}}}
      {p_N^{\,+}}
      {+\frac{{\mathbf q}_\perp}{2}}
      \equiv\left[\frac{M^2_N+{\mathbf q}_\perp^2/4}{p_N^{\,+}}\,,{\tilde p}'_N\right].
\label{eq:28}
\end{eqnarray}
With such a choice, the processes with vacuum pair production are suppressed, and the current  matrix elements can be computed as a simple overlap of Fock-space wavefunctions, with all off-diagonal terms involving pair production or annihilation by the current or vacuum vanishing. In the present meson-cloud model, we need to consider the contributions from the diagonal overlap between the bare-nucleon state, on one side, and the $BM$ components, on the other side. Furthermore, the electromagnetic current is a sum of one-body currents, $J^+=\sum_{B,M} J^+_B+J^+_M$,
 which involves individual hadrons one at a time. 
This corresponds  to assuming that there are no interactions among the particles 
in a multiparticle Fock state during the interaction with the photon. 
Therefore the external probe can scatter either on the bare nucleon,
$|N\rangle$, or one of the constituents  of the higher Fock states, $|BM\rangle$.
As a result, the matrix elements of the electromagnetic current can be written as the sum of the following two contributions
\begin{eqnarray}
& &\bra{\tilde p'_N,\lambda'_N,\tilde N} 
J^+_{V}\ket{\tilde p_N,\lambda_N,\tilde N}=
Z\,I^N_{\lambda'_N,\lambda_N}+
\delta I_{\lambda'_N,\lambda_N}.
\label{eq:current}
\end{eqnarray}
In Eq.~(\ref{eq:current}), $I^N$ is the contribution from the bare nucleon
corresponding to the diagram (a) in Fig.~\ref{fig:fig1}, and $\delta I$ 
is the contribution from the $BM$ Fock components of the physical nucleon.
This last term can further be split into two contributions, with the active
particle being the baryon ($\delta I^{(B'B)M}$) or the meson  
($\delta I^{(M'M)B}$), i.e.
\begin{eqnarray}
\delta I_{\lambda'_N,\lambda_N}=
\sum_{B,B',M} \,\delta I^{(B'B)M}_{\lambda'_N,\lambda_N}
+\sum_{M',M,B} \, \delta I^{(M'M)B}_{\lambda'_N,\lambda_N}.
\label{eq:ff_bm}
\end{eqnarray}

The $\delta I^{(B'B)M}_{\lambda'_N,\lambda_N}$ term in Eq.~(\ref{eq:ff_bm}) is schematically represented in Fig.~\ref{fig:fig1}(b) and is explicitly given by
\begin{eqnarray}
\delta I^{(B'B)M}_{\lambda'_N,\lambda_N}
&= &\sum_{B,B',M}
\sum_{\lambda,\lambda',\lambda''}
\int
{\rm d}y_B
\int
\frac{{\rm d}^2{\mathbf p}_{B\perp}}{2(2\pi)^3}
\bra{\tilde p_{B}+\tilde q,
\lambda',B'} 
J^+_B\ket{\tilde p_B,\lambda,B}\nn\\
& &\times
\phi^{\lambda_N\,(N,BM)}_{\lambda''\lambda}
(y_B,{\mathbf k}_{B\perp})\,
[\phi^{\lambda'_N\,(N,B'M)}_{\lambda''\lambda'}
(y_B,{\mathbf k}'_{B'\,\perp})]^*,
\label{eq:I_B}
\end{eqnarray}
where ${\mathbf k}_{B\perp}={\mathbf p}_{B\perp}-(1-y_B){\mathbf q}_{\perp}/2$ and
${\mathbf k'}_{B'\perp}={\mathbf p}_{B\perp}+(1-y_B){\mathbf q}_{\perp}/2$.

Analogously, the contribution from the meson in the $BM$ fluctuation 
is described by the diagram (c) in Fig. 1 and reads
\begin{eqnarray}
\delta I^{(M'M)B}_{\lambda'_N,\lambda_N}
& =&\sum_{B,M,M'}
\sum_{\lambda,\lambda',\lambda''}
\int
{\rm d}y_M
\int
\frac{{\rm d}^2
{\mathbf p}_{M\perp}}{2(2\pi)^3}
\bra{\tilde p_{M}+\tilde q,
\lambda',M'} J^+
\ket{\tilde p_M,\lambda,M}\nn\\
& &\times
\phi^{\lambda_N\,(N,BM)}_{\lambda''\lambda}(1-y_M,-{\mathbf k}_{M\perp})\,
[\phi^{\lambda'_N\,(N,BM')}_{\lambda''\lambda'}(1-y_M,-{\mathbf k'}_{M'\perp})]^*,
\label{eq:I_M}
\end{eqnarray}
with  ${\mathbf k}_{M\perp}={\mathbf p}_{M\perp}-(1-y_M){\mathbf q}_{\perp}/2$ and
${\mathbf k'}_{M'\perp}={\mathbf p}_{M\perp}+(1-y_M){\mathbf q}_{\perp}/2$.

As a result,  the contribution from the $BM$ components in Eqs.~(\ref{eq:I_B}) and (\ref{eq:I_M}) 
 are obtained by folding the current matrix elements of the baryon and meson constituents with the probability amplitudes describing the distributions of these constituents in the dressed initial and final nucleon. In general, the current matrix elements $\langle \tilde p+\tilde q| J^+|\tilde p\rangle$
appearing inside the integrals
 depend on the internal momentum of the baryon-meson state. However, 
as discussed in Ref.~\cite{Keister}, the kinematical nature of a light-front 
boost allows us to transform these matrix elements to a frame with 
$\vec p=0$, with the result
\bea
I_{\lambda'\lambda}(Q^2)
=\langle \tilde p+\tilde q, \lambda'|J^+|\tilde p,\lambda\rangle
=\langle M, {\mathbf q}_\perp,\lambda'|J^+|M, {\mathbf 0}_\perp,\lambda\rangle.
\eea
As a consequence, the current matrix elements  in Eqs.~(\ref{eq:I_B}) and (\ref{eq:I_M}) factor out of the internal momentum integration, and one finds
\bea
\delta I^{(B'B)M}_{\lambda'_N,\lambda_N}
&= &\sum_{B,B',M}
\sum_{\lambda,\lambda',\lambda''}
I^{B'B}_{\lambda'\lambda}(Q^2)\nn\\
& &\times
\int
{\rm d}y_B
\int
\frac{{\rm d}^2{\mathbf p}_{B\perp}}{2(2\pi)^3}
\phi^{\lambda_N\,(N,BM)}_{\lambda''\lambda}
(y_B,{\mathbf k}_{B\perp})\,
[\phi^{\lambda'_N\,(N,B'M)}_{\lambda''\lambda'}
(y_B,{\mathbf k}'_{B'\,\perp})]^*,
\label{eq:I_B2}
\\
\delta I^{(M'M)B}_{\lambda'_N,\lambda_N}
& =&\sum_{B,M,M'}
\sum_{\lambda,\lambda',\lambda''}
I^{M'M}_{\lambda'\lambda}(Q^2)\nn\\
& &\times
\int
{\rm d}y_M
\int
\frac{{\rm d}^2
{\mathbf p}_{M\perp}}{2(2\pi)^3}
\phi^{\lambda_N\,(N,BM)}_{\lambda''\lambda}(1-y_M,-{\mathbf k}_{M\perp})\,
[\phi^{\lambda'_N\,(N,BM')}_{\lambda''\lambda'}(1-y_M,-{\mathbf k}'_{M'\perp})]^*.\nn\\
& &
\label{eq:I_M2}
\end{eqnarray}
We also note that the sum in Eq.~(\ref{eq:ff_bm}) over all the possible $BM$ 
configurations leads to contributions from both the diagonal
 current matrix elements with the same hadrons in the initial and final state
($B'=B$  and $M'=M$ in Eq.~(\ref{eq:I_B2}) and (\ref{eq:I_M2}), respectively),
and the current matrix elements involving the electromagnetic transition
between different hadron states 
(i.e. the terms with $B'\neq B$ and $M'\neq M$ 
in Eq.~(\ref{eq:I_B2}) and (\ref{eq:I_M2}), respectively).

Finally, the current matrix element for the bare hadron states can be calculated as  overlap integrals of the hadron LCWF (see, e.g., Ref.~\cite{Brodsky:1989pv,cardarelli})
\bea
& &\bra{\tilde p'_{H'},\lambda'_{H'},H'} 
J^+_{V}\ket{\tilde p_H,\lambda_H,H}\nn\\
&&=
\sum_j\, e_j \sum_{\lambda_i}
\int[{\rm d}x][{\rm d}^2{\mathbf k}_\perp]
[\Psi_{\lambda'_{H'}}^{H',[f]}
(\{x_i,{\mathbf k}'_{\perp i};\lambda_i,\tau_i\}_{i=,1,..,N})]^*
\Psi_{\lambda_H}^{H,[f]}
(\{x_i,{\mathbf k}_{\perp i};\lambda_i,\tau_i\}_{i=,1,..,N}),\nn\\
& &
\eea
where $ {\mathbf k}'_{\perp j}={\mathbf k}_{\perp j}+(1-x_j){\mathbf q}_{\perp}$
for the struck quark, and  $ {\mathbf k}_{\perp i}=x_i{\mathbf q}_{\perp}$
for the spectator quarks.

The convolution formulae derived for the electromagnetic form factors can be extended to the calculation of the proton axial form factor $G_A(Q^2)$. In this case we need to consider
the diagonal matrix element of the axial current $J_A^+$, i.e.
\bea
G_A(Q^2)&=&
\langle \tilde p+\tilde q, \frac{1}{2}|J^+_A|\tilde p,\frac{1}{2}
\rangle.
\eea

The structure of Eqs.~(\ref{eq:current}), (\ref{eq:ff_bm}), (\ref{eq:I_B2}) and (\ref{eq:I_M2}) apply also  for the axial matrix elements, with the difference that the active mesons in the $\delta I^{(M'M)B}$ contribution can only be vector mesons. Furthermore, the LCWF overlap representation of the axial matrix elements between 
bare hadron states reads
\bea
& &\bra{\tilde p'_{H'},\lambda'_{H'},H'} 
J^+_{A}\ket{\tilde p_H,\lambda_H,H}\nn\\
&&=
\sum_j \tau_j  \sum_{\lambda_i} {\rm sign} (\lambda_j)
\int[{\rm d}x][{\rm d}^2{\mathbf k}_\perp]
[\Psi_{\lambda'_{H'}}^{H',[f]}
(\{x_i,{\mathbf k}'_{\perp i};\lambda_i,\tau_i\}_{i=,1,..,N})]^* 
\Psi_{\lambda_H}^{H,[f]}
(\{x_i,{\mathbf k}_{\perp i};\lambda_i,\tau_i\}_{i=,1,..,N}).\nn\\
& &
\eea



\section{Results and discussion}
\label{sect:results}

As we are interested in studying the effects of the meson cloud that notoriously manifest themselves at low values of $Q^2$, the three free parameters of the model, $\beta$, $\gamma$ in Eq.~(\ref{eq:76}) and the quark mass $m$, are fixed by fitting 8 experimental values of the proton and neutron form factors at low $Q^2$. In the fit procedure we have used the Sachs form factors  defined in terms of Dirac and Pauli form factors as
\be
G_E(Q^2) = F_1(Q^2) - \frac{Q^2}{4M_N^2} F_2(Q^2), \qquad G_M(Q^2) = F_1(Q^2) +F_2(Q^2).
\ee
The electric form factors are normalized as usual, i.e. $G_E^p(0) = 1$, $G_E^n(0) = 0$, and the magnetic form factors at $Q^2=0$ are normalized to the nucleon magnetic moments, i.e. $G_M^{p,n}(0) = \mu^{p,n}$. We have chosen to fit $\mu^p$, $\mu^n$, the proton axial coupling constant $g_A=G_A(0)$, $G_E^n$ at $Q^2=0.15$ GeV$^2$, and $G_E^p$ and $G_M^p$  at $Q^2=0.15$ and  $0.45$ GeV$^2$. 

A 5\% uncertainty was allowed in the fitting procedure. The multidimensional integration required for the numerical computation was implemented in a parallel computation using the parallelized version of the VEGAS routine of Ref.~\cite{Kreckel}.

The fitted values are $\gamma=3.21$, $\beta=0.489$ GeV and $m=0.264$ GeV.
These values differ from the original set of parameters in Ref.~\cite{Schlumpf94a}, used for the calculation of the nucleon electromagnetic form factors in a three-valence-quark model of the nucleon. They were  fitted only to the anomalous magnetic moments  of the proton and neutron to obtain
$\gamma=3.5$, $\beta=0.607$ GeV and $m=0.263$ GeV.

\begin{table}
\caption{\small Values of the form factors at $Q^2=0$ in the calculation of
 Ref.~\cite{Schlumpf94a} (first column), from the bare nucleon contribution
to the current matrix elements (second column), 
from the contribution of the baryon (third column) and meson (fourth column) 
state in the $BM$ component of the dressed nucleon.
The column labeled TOT gives the total result in the meson-cloud model, while in the last column (labeled Exp.)are given the experimental values.}
\label{table2} 
\begin{center}
\begin{tabular}{c|r|r|c|c|r|r}
\hline\hline
&
Ref.~\cite{Schlumpf94a}
&$Z\, I^N$
&$\sum_{B,B',M}\delta I^{(B'B)M}$ 
& $\sum_{M,M',B}\delta I^{(M'M)B}$ 
& TOT &Exp.~\cite{PDG}\\
\hline
$\mu^p$
&$2.78$
&$2.52$
&$\ \ 0.18$
&$\ \ 0.17$
& $2.87$
& $2.793$
\\
 $\mu^n$
&$-1.69$
&$-1.51$
&$-0.12$
&$-0.17$
&$-1.80$
& $-1.913$
\\
$g_A$
&$1.24$
&$1.12$
&$0.075$
&$0.002$
&$1.20$
&$1.2670$
\\
\hline\hline
\end{tabular}
\end{center}
\end{table}

In Table~\ref{table2} the values of $\mu^p$, $\mu^n$ and $g_A$ found in Ref.~\cite{Schlumpf94a} are compared with those obtained here and the experimental values~\cite{PDG}. As in Ref.~\cite{Schlumpf94a} and quite generally in the light-front formalism (see, e.g., Refs.~\cite{Dziemb88a, Chung91,cardarelli,DFS99,MaQS02}),  it is always difficult to reproduce all the three quantities simultaneously, so that some compromise has to be accepted. A significant improvement is obtained for $\mu^n$ by taking into account the meson cloud. One may appreciate that the two contributions in the baryon-meson fluctuation with the active particle being a meson ($\sum_{B,B',M}\delta I^{(B'B)M}$) or a baryon ($\sum_{M,M',B}\delta I^{(M'M)B}$) add up coherently in the right direction bringing the values of $\mu^p$, $\mu^n$ and $g_A$ closer to experiment.

\begin{figure}[ht]
\begin{center}
\epsfig{file=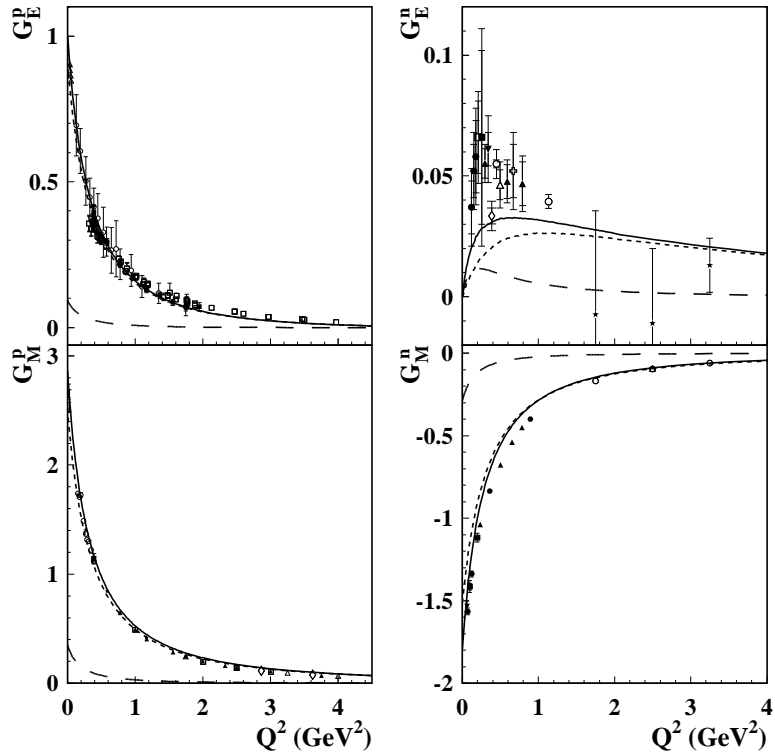,  width=12cm}
\end{center}
\caption{\small The four nucleon electromagnetic form factors compared with the world data considered in the analysis of Ref.~\cite{FW03} and the recent JLab data~\cite{Punjabi05} using $G_E^p=(\mu^p G_E^p/G_M^p)/(1+Q^2/0.71{ \rm GeV}^2)^2$ (open squares). Dotted (dashed) line for the contribution of the meson cloud (valence quarks). Solid line for the total result. }
\label{fig:fig2}
\end{figure}

The resulting electromagnetic form factors of both the proton and neutron are shown in Fig.~\ref{fig:fig2} in comparison with  the world data considered in the analysis of Ref.~\cite{FW03} and the recent JLab data~\cite{Punjabi05}. A rather good fit is obtained in the proton case in the whole range of available data, while in the neutron case the fit is less satisfactory. In any case, the contribution from the meson cloud is smooth and mainly significant for $Q^2< 0.5$ GeV$^2$ with a maximum at $Q^2=0$. Therefore, in agreement with dispersion relation analyses~\cite{Hammer04a, Hammer04b,BHM07} this model is unable to produce the bump/dip structure advocated in Ref.~\cite{FW03} around $Q^2=0.2-0.3$ GeV$^2$.
At higher $Q^2$ the explicit meson cloud contribution dies out, but indirectly
affects the bare-nucleon contribution through the normalization factor $Z$~\cite{Schutte}, which is equal to $0.91$ in our calculation. 

\begin{figure}[ht]
\begin{center}
\epsfig{file=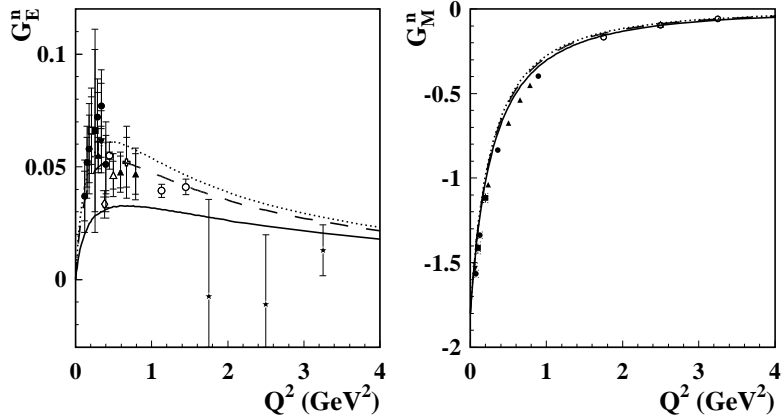,  width=12cm}
\end{center}
\caption{\small The electric form factor of the neutron. Data points and solid line as in Fig.~\ref{fig:fig2}. Dashed (dotted) line with 1\% (2\%) mixed-symmetry  $S'$-state in the bare neutron wavefunction.}
\label{fig:fig3}
\end{figure}

The results plotted in Fig.~\ref{fig:fig2} are all obtained starting from an instant-form wavefunction $\Psi_\lambda^{H,[c]}$ in Eq.~(\ref{eq:75}) containing a totally symmetric $S$-wave part in the quark momenta. In order to improve the result for the neutron electric form factor it is known that the presence of a small admixture (1--2\%) of mixed-symmetry $S'$-wave components is most important~\cite{CardSim00,Graz01,Graz02,Julia04,Julia06a}. Following~\cite{Julia04} we assume the mixed-symmetry $S'$-wave component to be represented by appropriate combination of mixed-symmetry spin-isospin wavefunctions with two radial wavefunctions of mixed symmetry of the form
\be
\Psi^{N,[c]}_s({\mathbf p},{\mathbf q}) = {\cal N}_s\frac{p^2-q^2}{p^2+q^2} \Psi^{N,[c]}(\{{\mathbf k}_i\}_{i=1,3}), \quad \Psi^{N,[c]}_a ({\mathbf p},{\mathbf q}) = {\cal N}_a\frac{{\mathbf p}\cdot{\mathbf q}}{p^2+q^2}\Psi^{N,[c]}(\{{\mathbf k}_i\}_{i=1,3}), 
\ee
where $\Psi^{N,[c]}(\{{\mathbf k}_i\}_{i=1,3})$ is the symmetric $S$-wavefunction (\ref{eq:76}), ${\cal N}_s$ and ${\cal N}_a$ are normalization factors, and $\mathbf p$ and $\mathbf q$ are the Jacobi coordinates
\be
{\mathbf p} = -\sqrt{\frac{3}{2}}({\mathbf k}_1 +{\mathbf k}_2), \quad {\mathbf q} = \sqrt{\frac{1}{2}}({\mathbf k}_1 -{\mathbf k}_2).
\ee
The consequences of including a small percentage of such mixed-symmetric contribution to the neutron electric form factor are illustrated in Fig.~\ref{fig:fig3}. Even a percentage as small as 1\% is able to produce a quite good result compared to data. As anticipated in Ref.~\cite{Julia04} the same calculation leaves the other nucleon form factors almost unaffected.

\begin{figure}[ht]
\begin{center}
\epsfig{file=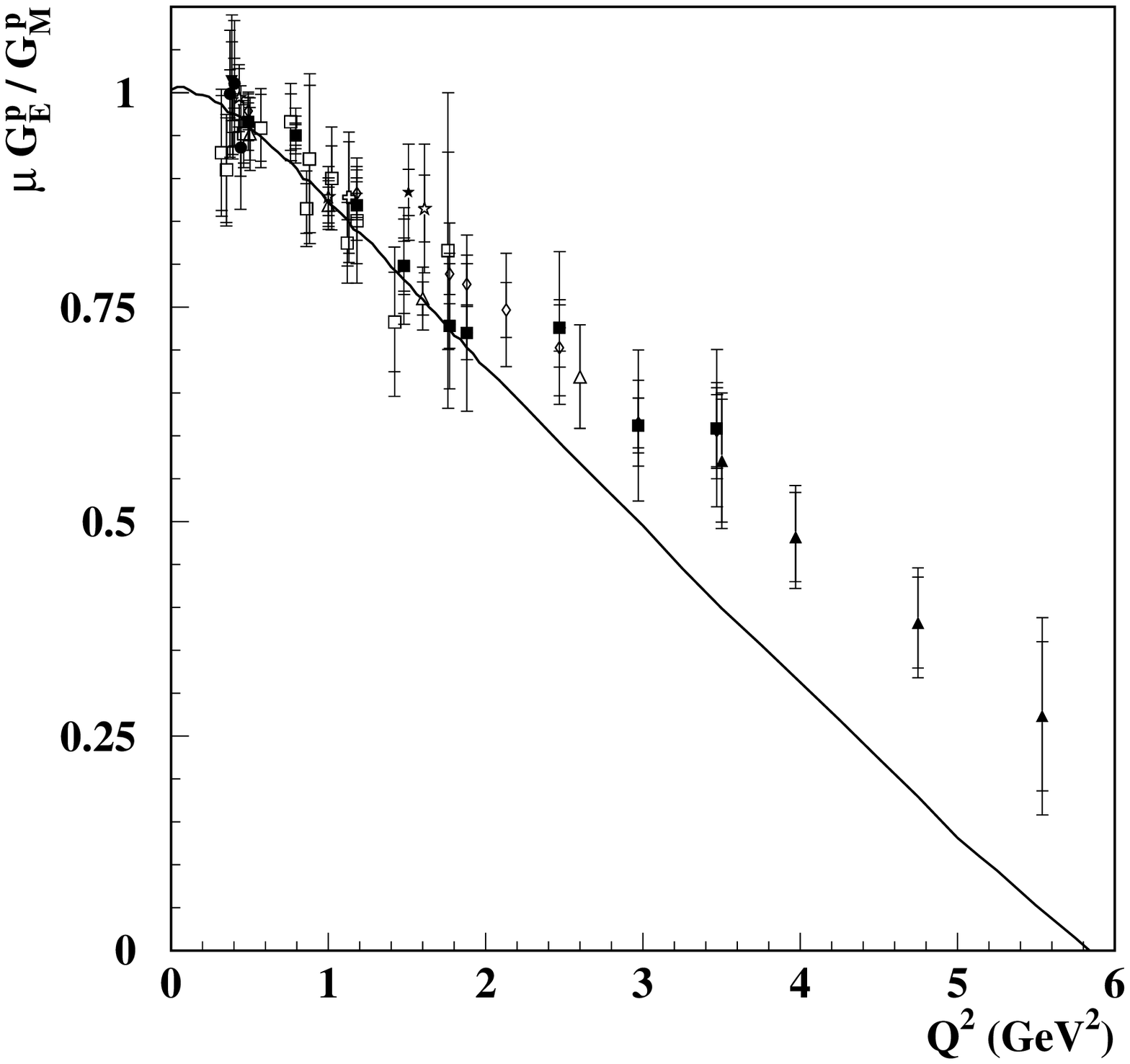,  width=12cm}
\end{center}
\caption{\small The electric to magnetic form factor ratio $\mu^p G_E^p/G_M^p$ for the proton. Data points as in Fig.~\ref{fig:fig2}.}
\label{fig:fig4}
\end{figure}

The slope of the electric form factor at $Q^2=0$ determines the nucleon charge radius, i.e.
\be
r^2_{p,n} 
= {}- 6\left. \frac{{\rm d}G_E^{p,n}(Q^2)}{{\rm d}Q^2}\right\vert_{Q^2=0}.
\ee
The corresponding values for proton and neutron obtained using an SU(6) symmetric or a mixed-symmetric instant-form wavefunction are reported in Table~\ref{table3}, where also the partial contributions are indicated when one considers either the bare nucleon or the contribution of the cloud with an active baryon or meson. Rather good values of $r_p$ and $r_n$ are obtained in the latter case. For the proton the charge radius is mostly due to the valence quarks, i.e. the bare proton. The meson cloud brings a contribution of about 5\% which leads to a final value of $0.877$ fm, in close agreement with the experimental value $0.8750\pm 0.0068$~\cite{PDG}.  Including the mixed--symmetry $S'$-wave component the charge radius of the bare proton is 0.837 fm. The cloud adds a small contribution, that makes the total value to be slightly overestimated. In the neutron case the bare contribution is quite small, as expected. In contrast, both the cloud and the mixed symmetry are equally important. The contribution of the active meson in the cloud is substantial and with the right sign. Including also the mixed-symmetry $S'$-wave component $r^2_n$ becomes quite close to the experimental value $-0.1161\pm 0.0022$~\cite{PDG}.

\begin{table}
\caption{\small The different contributions (in fm$^2$) to the proton and neutron mean square charge radii, $r^2_p$ and $r^2_n$ respectively,  from the bare nucleon and when the baryon or the meson is active in the cloud. The column labeled TOT is the total result. The lines labeled SU(6) (mixed symmetry) refer to the symmetry of the instant-form wavefunction of the bare nucleon.}
\label{table3} 
\begin{center}
\begin{tabular}{c|c|c|c|c|c} 
\hline\hline
& & bare nucleon & active baryon & active meson & TOT \\  
\hline
$r^2_p$\quad{} & SU(6) & $0.64$ & $0.065$  & $0.061$ & $0.77$ \\ 
$r^2_p$\quad{} & mixed symmetry & $0.70$& $0.065$ & $0.061$ & $0.82$ \\ 
$r^2_n$\quad{} & SU(6) & $-0.0097$ & $0.0085$ & $-0.063$ & $-0.064$ \\ 
$r^2_n$\quad{} & mixed symmetry & $ -0.058$ & $0.0085$ & $-0.063$ & $-0.112$ \\ 
\hline\hline
\end{tabular}
\end{center}
\end{table}

In Fig.~\ref{fig:fig4} the electric to magnetic form factor ratio $\mu^p G_E^p/G_M^p$ is shown for the proton. The model follows the observed trend of data taken in polarized elastic electron scattering with a steepest fall-off at values of $Q^2$ much larger than those involved in the fit to determine the model parameters. This is due to the combined effect of a slightly overestimated $G_M^p$  and a slightly underestimated $G_E^p$ at large values of $Q^2$ (see Fig.~\ref{fig:fig2}), where in practice only the valence quarks contribute. The result is only slightly modified when including a mixed-symmetry contribution in the nucleon wavefunction.

\begin{figure}[ht]
\begin{center}
\epsfig{file=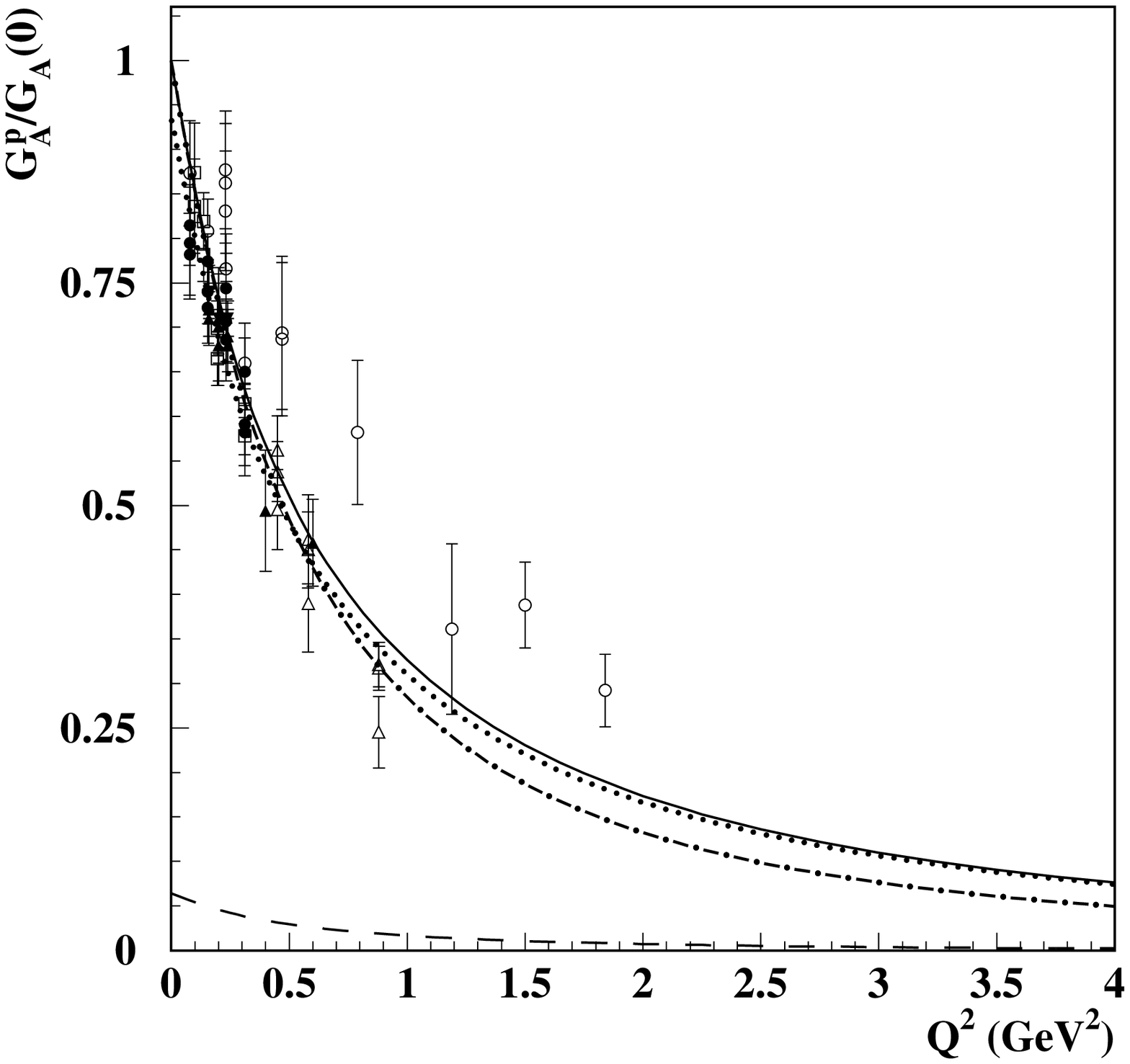,  width=12cm}
\end{center}
\caption{\small The axial  form factor of the proton. Dotted (dashed) line for the contribution of the meson cloud (valence quarks). Solid line for the total result. Dot-dashed line for the phenomenological dipole form. Data points are the world data considered in Ref.~\cite{BEM02}.}
\label{fig:fig5}
\end{figure}

The predicted axial form factor of the proton is shown in Fig.~\ref{fig:fig5}. The axial form factor has been normalized by its value at $Q^2=0$, i.e. by the fitted value of the axial coupling constant $g_A$. Also in this case the meson-cloud contribution is only significant at low values of $Q^2$, although not sufficient to bring $g_A$ in complete agreement with experiment. However, the observed dipole form of the axial form factor, i.e. $G_A(Q^2)/G_A(0)=1/(1+Q^2/M_A^2)^2$ with $M_A =1.069$ GeV, is well reproduced. 

\begin{figure}[ht]
\begin{center}
\epsfig{file=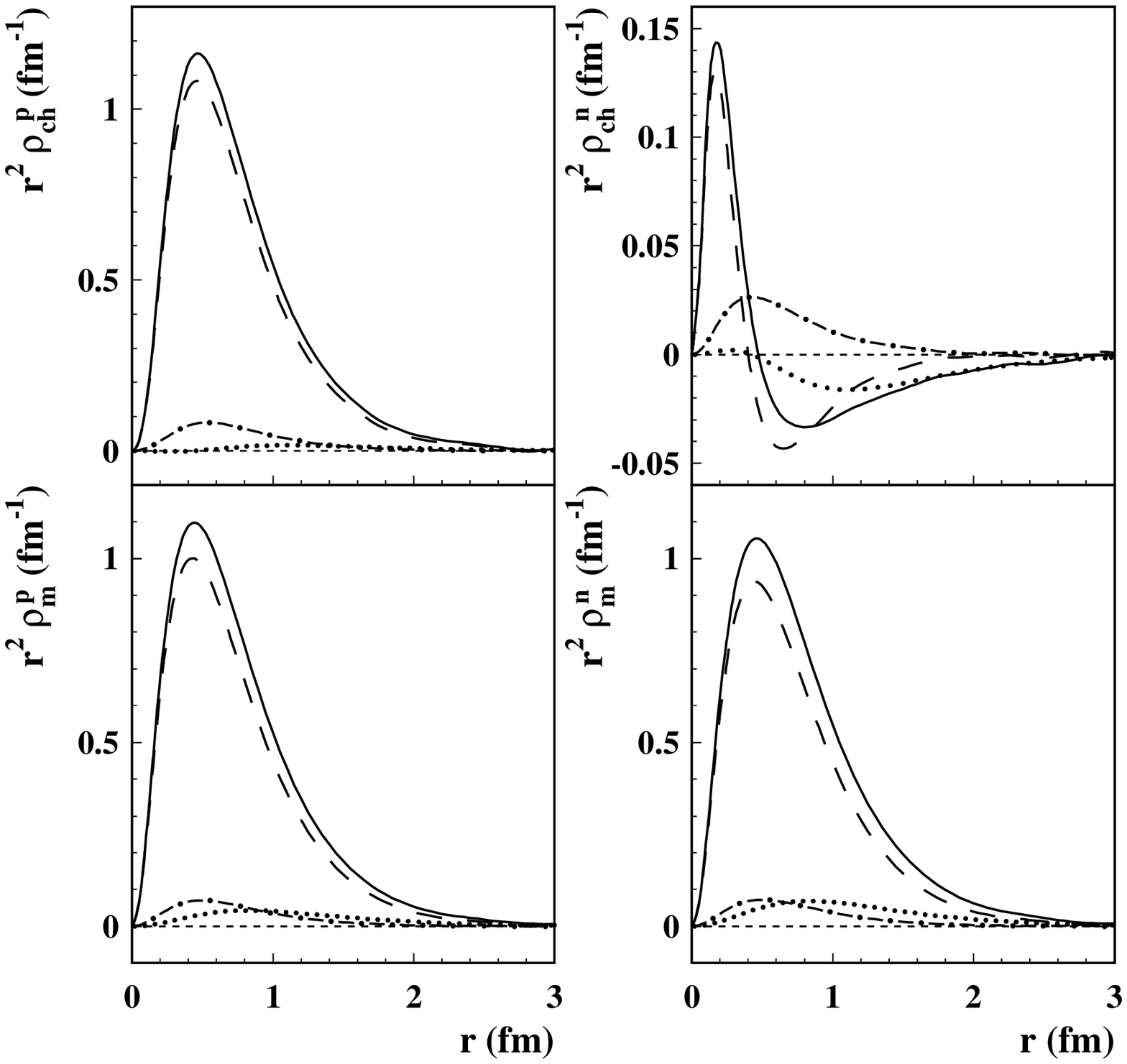,  width=12cm}
\end{center}
\caption{\small The proton and neutron charge and magnetization densities. Dashed, dot-dashed and dotted lines for contributions from the bare nucleon, the active baryon and the active meson in the cloud. Solid line for the total result.}
\label{fig:fig6}
\end{figure}

Neglecting relativistic corrections, in the Breit frame the radial distribution of the nucleon charge ($\rho_{ch}$) and magnetization ($\rho_m$) are given by the Fourier-Bessel transform of the nucleon electromagnetic Sachs form factors, i.e.
\bea
\rho^{p,n} _{ch}(r) & = & \frac{2}{\pi}\int {\rm d}Q\, Q^2 \,j_0(Qr) \, 
G_{E}^{p,n}(Q^2),\\
\mu^{p,n}\rho^{p,n} _{m}(r) & = & \frac{2}{\pi}\int {\rm d}Q\, Q^2 \,j_0(Qr) \, G_{M}^{p,n}(Q^2).
\eea
The corresponding results are shown in Fig.~\ref{fig:fig6} together with the partial contributions from the bare nucleon and when the baryon or the meson is active in the cloud. In all cases the meson-cloud contribution is rather smooth and dies out beyond 2 fm. With the exception of the neutron charge density, when the baryon is active its contribution is peaked at approximately the same position as the bare-nucleon contribution, while the active meson is peaked at $\sim 1$ fm. Thus the meson cloud manifests itself as a slight extension of the radial distribution up to $\sim 2 $ fm as suggested by the analysis of Ref.~\cite{FW03}. In the case of the neutron charge distribution the two components of the meson cloud behave differently. The active baryon, a proton or a $\Delta^+$, gives a positive contribution of the same shape as the bare proton and is appropriately scaled by the corresponding vertex functions. The contribution of the active meson, a $\pi^-$, is opposite and peaked at $\sim 1.3$ fm. The resulting charge distribution shows a positive core surrounded by a negative surface charge pushed outwards by the meson cloud  and peaking at $\sim 0.8$ fm, in agreement with the analysis of Ref.~\cite{Kelly02} and the expectation based on the picture of a hadron's periphery caused by the pion cloud~\cite{Lvov}.

\begin{figure}[ht]
\begin{center}
\epsfig{file=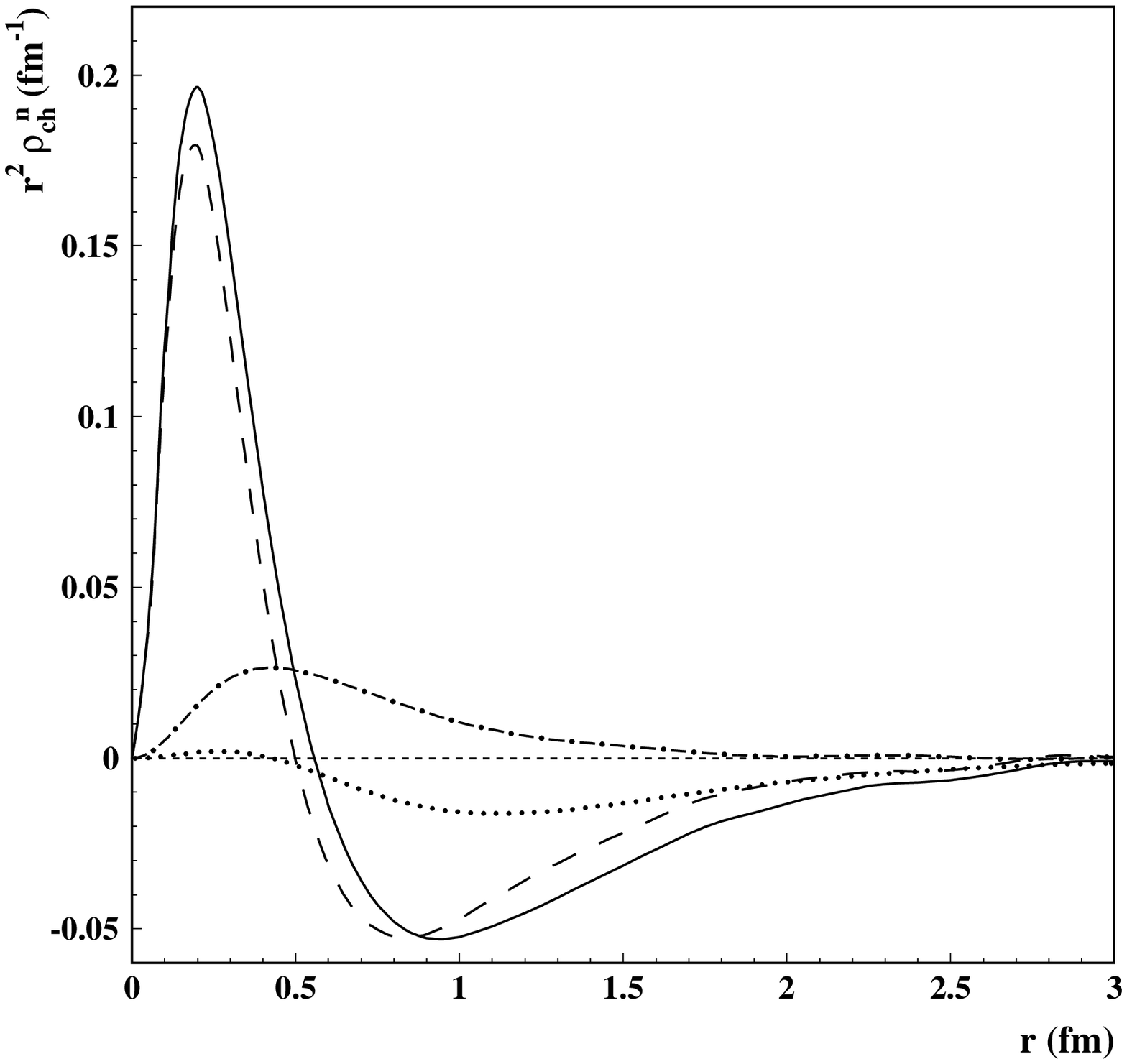,  width=12cm}
\end{center}
\caption{\small The neutron charge density with a mixed-symmetry $S'$-wave component in the neutron wavefunction. Dashed, dot-dashed and dotted lines for contributions from the bare nucleon, the active baryon and the active meson in the cloud. Solid line for the total result.}
\label{fig:fig7}
\end{figure}

The effect of including the mixed-symmetry $S'$-wave component in the neutron case can be appreciated from Fig.~\ref{fig:fig7}. The inner  positive core is more pronounced and the negative surface charge is even more pushed outwards. Consequently, the (negative) mean square radius approaches the experimental value (see Table~\ref{table3}).

The direct relationship between Sachs form factors and the static charge and magnetization densities is lost when relativity is considered because there is a variation with $Q^2$ of the Breit frame and electron scattering measures transitions matrix elements between nucleon states that have different momenta. Therefore one has to apply appropriate boosts that in a relativistic composite system such as the nucleon depend on the interaction among its constituents. The problem of finding a suitable prescription to relate Sachs form factors to the static charge and magnetization densities was recently addressed in Ref.~\cite{Kelly02} taking into account the Lorentz contraction of the densities in the Breit frame relative to the rest frame. The consequences are that relativity tends to pull the density inward and to amplify oscillations at large radii. This has been confirmed in the model of Ref.~\cite{FGLP06}. The same effect should be expected also here. In any case in the present analysis the meson cloud is responsible for a long-range contribution to the nucleon charge and magnetization densities.

\begin{figure}[ht]
\begin{center}
\epsfig{file=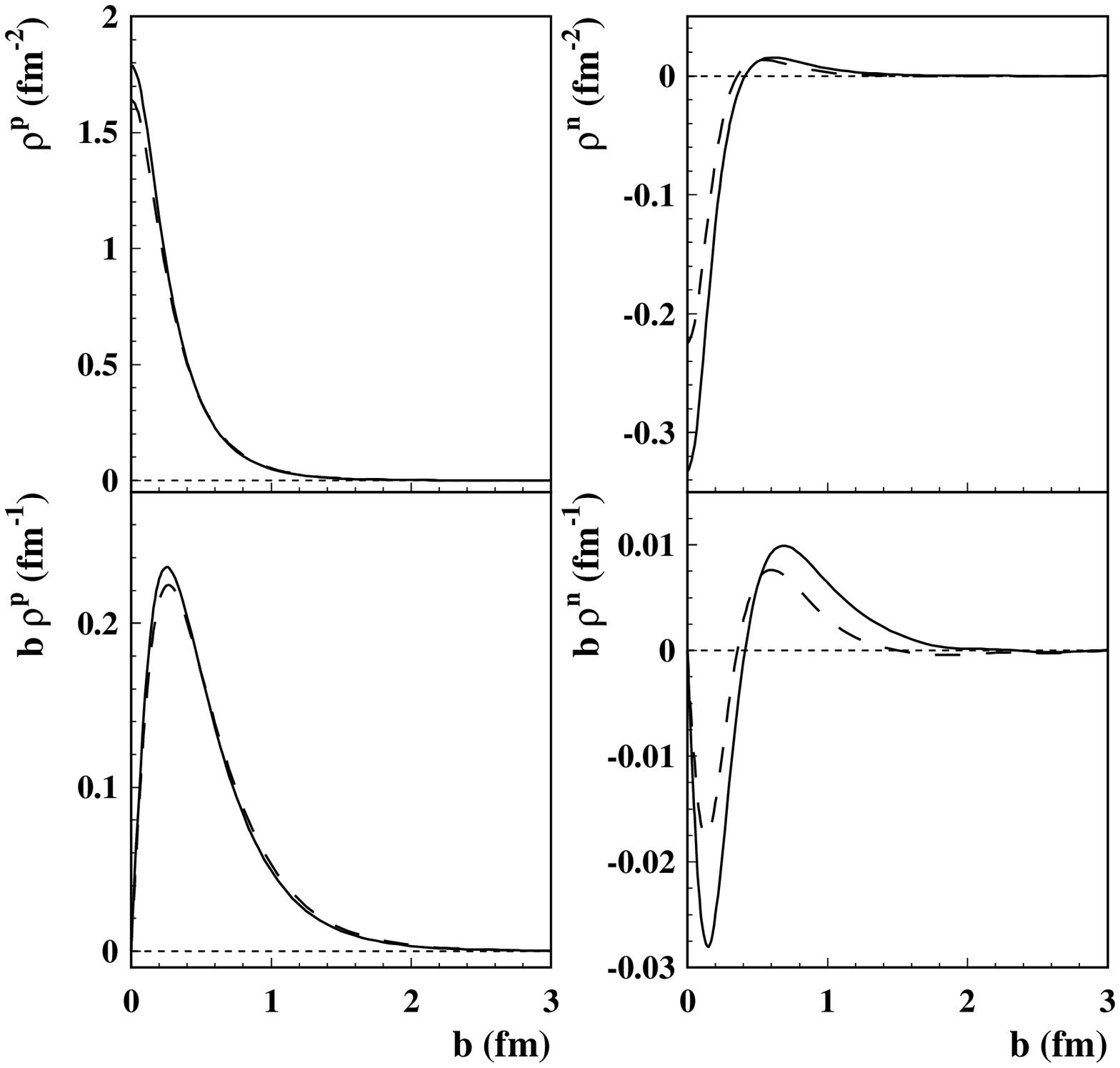,  width=12cm}
\end{center}
\caption{\small The proton and neutron charge density as a function of the impact parameter $b$. Solid lines for a permutationally symmetric momentum wavefunction, dashed lines with mixed-symmetry components included.}
\label{fig:fig8}
\end{figure}

The problem of unambiguously determining the charge density can be solved by looking at the charge density $\rho(b)$ of partons in the transverse (impact parameter) plane with respect to the direction of the three-momentum transfer~\cite{Miller07}. This is possible because in the transverse plane boosts are purely kinematical, i.e. in the light-front framework they form a Galilei subgroup of the Poincar\'e group~\cite{KS70,Burkardt02}. Then $\rho(b)$ is the two-dimensional Fourier transform of the Dirac form factor $F_1$:
\be
\label{eq:rhodib}
\rho(b) = \frac{1}{2\pi}\int_0^\infty {\rm d}Q\,Q \,J_0(Qb) F_1(Q^2),
\ee
where $J_0 $ is a cylindrical Bessel function. 

The corresponding charge densities for the proton and the neutron are plotted in Fig.~\ref{fig:fig8}. As in Ref.~\cite{Miller07} the densities are concentrated at low values of $b$ with a positive peak for the proton and a negative peak for the neutron. 

These nucleon charge densities can be related to quark transverse distributions. Assuming that only up and down quarks are in the nucleon and invoking isospin symmetry, we have
\bea
\rho^p(b) & = & \frac{4}{3} u(b) - \frac{1}{3} d(b), \\
\rho^n(b) & = & -\frac{2}{3} u(b) + \frac{2}{3} d(b),
\eea
where $u(b)$ is the transverse distribution for an up quark in the proton or a down quark in the neutron, and $d(b)$ is the transverse distribution for a down quark in the proton or an up quark in the neutron. Both $u(b)$ and $d(b)$ are normalized to 1. They can be obtained using
\bea
u(b) & = &  \rho^p(b) +\frac{1}{2} \rho^n(b), \\
d(b) & = & \rho^p(b) + 2\rho^n(b).
\eea

\begin{figure}[ht]
\begin{center}
\epsfig{file=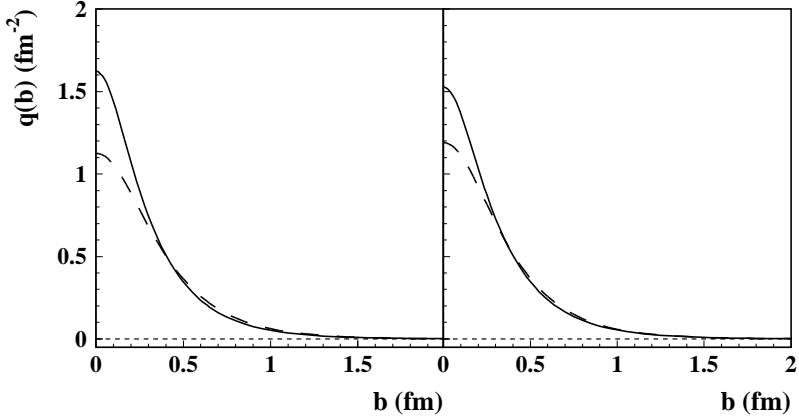,  width=12cm}
\end{center}
\caption{\small Transverse distributions of up (solid lines) and down (dashed lines) quarks in the proton as a function of the impact parameter $b$ with a permutationally symmetric momentum wavefunction of the bare nucleon (left panel), and with a mixed-symmetric component (right panel).}
\label{fig:fig9}
\end{figure}

The resulting distributions are shown in Fig.~\ref{fig:fig9} in the two cases of a permutationally symmetric momentum wavefunction of the bare nucleon and of an included mixed-symmetric component. The central up quark density turns out to be larger than that of the down quark by about 40\% in the symmetric case and by about 25\% including the mixed symmetry. Quite similar results (about 30\%) have been obtained in Ref.~\cite{Miller07} using phenomenological parametrizations of the Sachs form factors and deducing $F_1$ in terms of $G_E$ and $G_M$.

\begin{figure}[ht]
\begin{center}
\epsfig{file=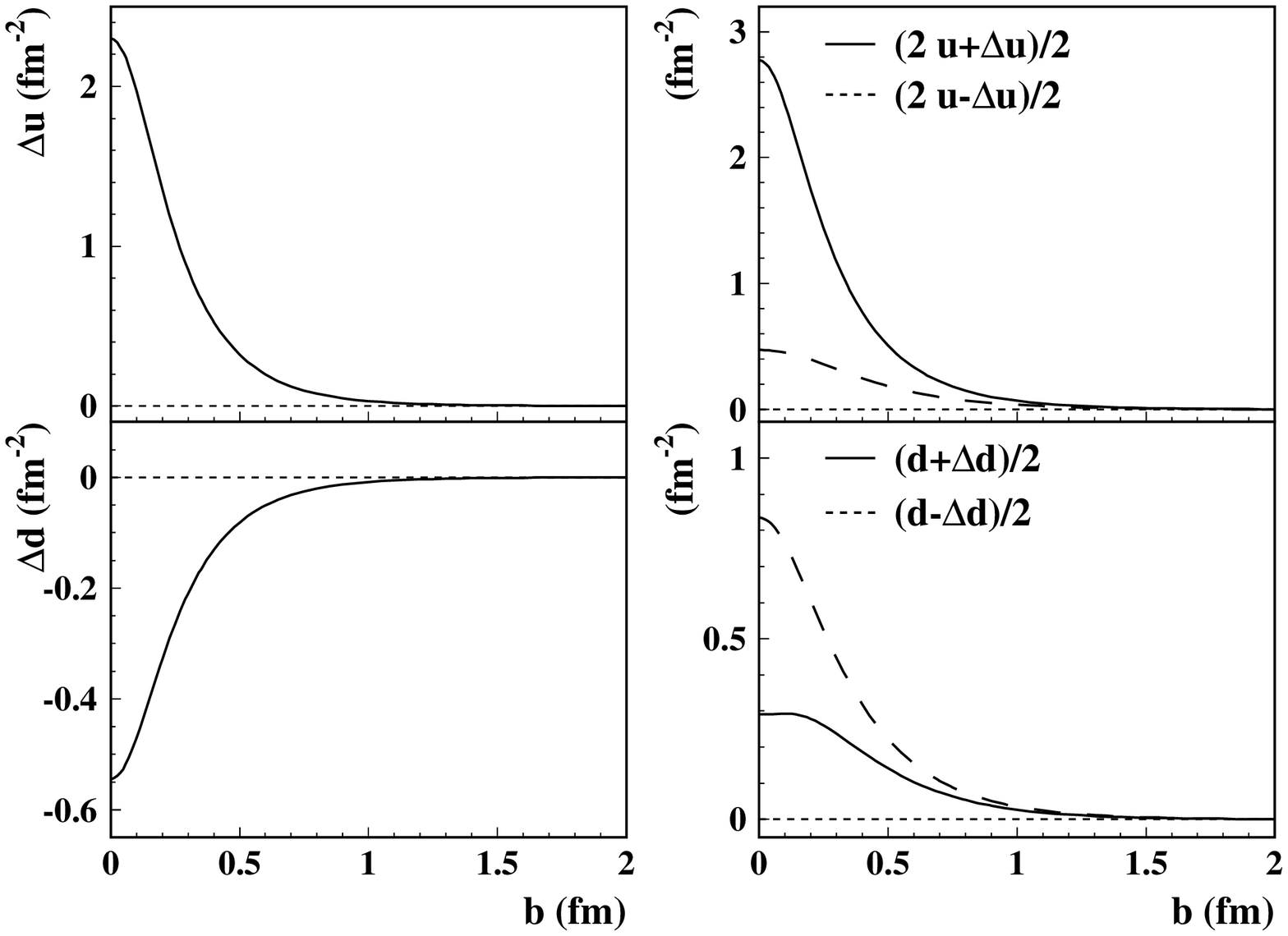,  width=12cm}
\end{center}
\caption{\small Transverse distribution of up and down quarks in a longitudinally polarized proton as a function of the impact parameter $b$. Left panels: the axial contributions $\Delta u$ and $\Delta d$ for up and down quarks, respectively. Right panels: total contribution for quarks polarized in the longitudinal direction, either parallel (solid lines) or antiparallel (dashed lines) to the proton helicity.}
\label{fig:fig10}
\end{figure}

The probability $\rho^q(b,\lambda,\Lambda)$ to find a quark with transverse position $b$ and light-cone helicity $\lambda$ ($=\pm 1$) in the nucleon with longitudinal polarization $\Lambda$ ($=\pm 1$) can be expressed as Fourier transform of the combination of the quark contributions to the Dirac and axial form factors~\cite{PB07}, i.e.
\bea
\rho^q(b,\lambda,\Lambda) &=& \frac{1}{2}\int {\rm d}^2\mathbf{q}_\perp
\left[ F^q_1(Q^2=\mathbf{q}^2_\perp) +\lambda\Lambda G^q_A(Q^2=\mathbf{q}^2_\perp) \right]\,e^{i\mathbf{q}_\perp\cdot\mathbf{b}} \nonumber \\
&=& \frac{1}{4\pi}\int {\rm d}Q \,Q J_0(Qb) \left[ F^q_1(Q^2) + \lambda\Lambda G^q_A(Q^2) \right] \nonumber \\
&\equiv& \frac{1}{2}\left[\rho^q(b) +  \lambda\Lambda \Delta q(b)\right],
\eea
where $\rho^q(b)$ was already defined in Eq.~(\ref{eq:rhodib}) and $\Delta q(b)$ is the Fourier transform of $G^q_A(Q^2)$. Assuming a positive proton helicity ($\Lambda=1$) the resulting probability is shown in the right panels of Fig.~\ref{fig:fig10}. The axial contributions $\Delta u (b)$ and $\Delta d (b)$ for up and down quarks (left panels), respectively, have opposite sign. When suitably combined with the corresponding transverse distributions $u(b)$ and $d(b)$ given in Fig.~\ref{fig:fig9} we see that the positive helicity up quarks in the proton are preferentially aligned with the proton helicity, while the opposite occurs for down quarks. This result is in total agreement with that shown in  Fig. 7 of Ref.~\cite{PB07} where quite a different radial distribution of the axially symmetric spin density was presented for up and down quarks in  the transverse plane.


\section{Concluding remarks}
\label{sect:conclusion}

The meson-cloud model, as revisited in Ref.~\cite{PB06} to study generalized parton distributions and including baryon-meson configurations with the baryon being a nucleon or a $\Delta$ and the meson being a pion as well as a vector meson such as the $\rho$ or the $\omega$, has been used to describe the electroweak structure of the nucleon. Light-cone wavefunctions for the bare nucleon were constructed starting from the momentum wavefunction (\ref{eq:76}) taken from Ref.~\cite{Schlumpf94a} and depending on three parameters, the scale $\beta$, the parameter $\gamma$ for the power-law behaviour, and the quark mass $m$. They are determined by fitting 8 experimental values of the proton and neutron form factors at low $Q^2$. No other free parameters enter the model calculations, since all other ingredients are fixed from the beginning on the basis of previous analysis. 

An overall good fit to the electromagnetic form factors is obtained, with the exception of the neutron electric form factor where it is essential to also include a mixed-symmetry $S'$-wave momentum component, in agreement with earlier findings~\cite{CardSim00,Graz01,Graz02,Julia04,Julia06a}. This component only slightly affect the other form factors. In any case the meson-cloud contribution is smooth and only significant below $Q^2=0.5$ GeV$^2$. Therefore, as in analyses based on dispersion relations~\cite{Hammer04a, Hammer04b,BHM07}, also in this model no possibility exists to reproduce the bump/dip structure discussed in Ref.~\cite{FW03}. A similar smooth contribution arises in the proton axial form factor.

Charge and magnetization densities are deduced as a function of both the radial distance from the nucleon center and the transverse distance (impact parameter) with respect to the direction of the three-momentum transfer. 

The meson cloud produces a slight extension of the radial distribution of the static charge and magnetization up to $\sim 2$ fm. It is confirmed that the neutron charge distribution shows a positive core surrounded by a negative surface charge~\cite{Kelly02,Miller07}. It is pushed outwards by the long-range meson cloud with opposite contributions from the active baryon and meson in the baryon-meson  component of the nucleon wave function.

As a function of the impact parameter a central negative charge is found for the neutron.  A similar result has been obtained in Ref.~\cite{Miller07} starting from a phenomenological fit of the electromagnetic form factors. This result can be explained invoking isospin symmetry and observing that the up quark transverse distribution in the proton is larger than the down quark one, a result consistent with deep inelastic scattering data.

The up and down quark distributions associated with the Fourier transform of the axial form factor have opposite sign, with the consequence that the probability to find an up (down) quark with positive helicity is maximal when it is (anti)aligned with the proton helicity, in close agreement with the radial distribution of the axially symmetric spin density studied in Ref.~\cite{PB07}.

In conclusion, the meson-cloud model appears to give a satisfactory description of the electroweak properties of the nucleon with interesting information about its structure in the nonperturbative regime of QCD.


\section*{Acknowledgments}

This research is part of the EU Integrated Infrastructure Initiative Hadron Physics Project under contract number RII3-CT-2004-506078. The diagrams in the paper have been drawn using the Jaxodraw package~\cite{binosi}.


\appendix
\setcounter{section}{0}
\setcounter{equation}{0}
\renewcommand{\theequation}{\Alph{section}.\arabic{equation}}


\section{Vertex functions}
\label{appendixa}

In this Appendix we work out the case of the
$B\rightarrow B' V$ transitions in the light-front formalism, with
the baryon states $B$ being a nucleon or a $\Delta$, and the vector mesons 
$V$ corresponding to $\omega$, or $\rho$.
The vertex functions for the coupling of baryons with pseudoscalar mesons 
are given in Appendix C of Ref.~\cite{PB06}.
The  vertex functions for transitions to vector mesons can be found in several 
places~(see, e.g., Refs.\cite{HSS96, DHSS97,Speth98}), and
the longitudinal-momentum distributions corresponding to the 
integration over the transverse momentum of the squared vertex functions 
are explicitly given in Refs.~\cite{Cao,Fries,Kumano}.
The quoted results are controversial, in the sense that 
they differ although the formalism is exactly the same.
In particular, we agree with the results for the longitudinal-momentum 
distributions of Refs.~\cite{Fries,Kumano}, and differ from Ref.~\cite{Cao}, 
while for the vertex functions we agree with the conclusions drawn in the 
Appendix of Ref.~\cite{Kumano}, where the origin of the differences 
from Refs.\cite{HSS96, DHSS97,Speth98} is explained in 
details. However, Kumano et al.~\cite{Kumano} do not give explicit analytical 
expressions for the vertex functions, and we find convenient 
 to show here their derivation and in particular their dependence
on the transverse momentum which enters in the
convolution formulae for the form factors.

The light-front vectors are defined as
\begin{equation}
A^\mu=(A^-, A^+,{\mathbf A}_\perp),
\end{equation}
with
\begin{equation}
A^\pm=A^0\pm A^3,\quad {\mathbf A}_\perp=(A^1,A^2).
\end{equation}
We also use the notations $A_{R,L}=A^1\pm iA^2$ and $\tilde A=(A^+,{\mathbf A}_\perp)$.

The light-front nucleon spinors $u_\lambda(\tilde p)$ are 
given by
\begin{equation}
u_{1/2}(\tilde p)=\frac{1}{\sqrt{2p^+}}
\left(
\begin{array}{c}
p^+ + m\\
p_R\\
p^+-m\\
p_R
\end{array}
\right),
\qquad
u_{-1/2}(\tilde p)=\frac{1}{\sqrt{2p^+}}
\left(
\begin{array}{c}
-p_L\\
p^+ + m\\
p_L\\
m-p^+
\end{array}
\right).
\label{eq:spinor}
\end{equation}

The gamma matrices are defined as in Ref.~\cite{Bjo65}.

A similar expansion for the $\Delta$ field involves the Rarita-Schwinger 
spinors given by
\begin{eqnarray}
u^\mu_{3/2}(\tilde p)&=&\epsilon^\mu_{+1}(\tilde p)\,u_{1/2}(\tilde p),\nn\\
u^\mu_{1/2}(\tilde p)&=&\sqrt{\frac{2}{3}}\epsilon^\mu_{0}(\tilde p)\,u_{1/2}
(\tilde p)
+\sqrt{\frac{1}{3}}\epsilon^\mu_{+1}(\tilde p)\,u_{-1/2}(\tilde p)
,\nn\\
u^\mu_{-1/2}(\tilde p)&=&\sqrt{\frac{2}{3}}\epsilon^\mu_{0}(\tilde p)\,u_{-1/2}
(\tilde p)
+\sqrt{\frac{1}{3}}\epsilon^\mu_{-1}(\tilde p)\,u_{1/2}(\tilde p)
,\nn\\
u^\mu_{-3/2}(\tilde p)&=&\epsilon^\mu_{-1}(\tilde p)\,u_{-1/2}(\tilde p),
\end{eqnarray}
where the polarization vectors are given by
\begin{eqnarray}
\epsilon^\mu_{+1}(\tilde p)&=&\left(-\sqrt{2}\frac{p_R}{p^+},0,
(-\frac{1}{\sqrt{2}},
-\frac{i}{\sqrt{2}})\right),\nn\\
\epsilon^\mu_{0}(\tilde p)&=&\frac{1}{m}\left(\frac{{\mathbf p}_\perp^2-m^2}{p^+},
p^+,
{\mathbf p}_\perp\right),\nn\\
\epsilon^\mu_{-1}(\tilde p)&=&
\left(\sqrt{2}\frac{p_L}{p^+},0,(\frac{1}{\sqrt{2}},
-\frac{i}{\sqrt{2}})\right).
\end{eqnarray}

\begin{table}
\caption{\small Vertex functions for $N\to N'V$ and particle helicities $\oneh\to\lambda'_N,\lambda_V$ in the prescription A.}
\label{table4} 
\begin{center}
\begin{tabular}{r r c l}
\hline\hline
$\lambda'_N$ &  $\lambda_V$ & & $V(N,NV$)\\
\hline 
$+\oneh$&$+1$ & &$\frac{\sqrt{2}k_L}{\sqrt{y}}\left[\frac{g}{1-y}+\frac{f}{2}\right]$\\
\\
$-\oneh$&$+1$& &$g\frac{\sqrt{2}(M_B-yM_N)}{\sqrt{y}}-\frac{f}{2M_N}
\frac{\sqrt{2}[yM^2_V-(1-y)^2M_NM_B]}{\sqrt{y}(1-y)}$\\
\\
$\oneh$&$0$ & &$g\frac{(1-y)^2M_NM_B-yM_V^2+k^2_\perp}{M_V\sqrt{y}(1-y)}-\frac{f}{2M_N}\frac{M_V(yM_N-M_B)} {\sqrt{y}}$\\
\\
$-\oneh$&$0$& &$g\frac{k_R(M_B-M_N)}{M_V\sqrt{y}}-\frac{f}{2M_N}\frac{k_RM_V(1+y)}
{\sqrt{y}(1-y)}$\\
\\
$+\oneh$&$-1$& &$-g\frac{\sqrt{2}yk_R}{\sqrt{y}(1-y)}+\frac{f}{2M_N}\frac{\sqrt{2}k_R
M_B}{\sqrt{y}}$\\
\\
$-\oneh$&$-1$& &$-\frac{f}{2M_N}\frac{\sqrt{2}k^2_R}{\sqrt{y}(1-y)}$\\
\hline\hline
\end{tabular}
\end{center}
\end{table}

\begin{table}
\caption{\small Vertex functions for $N\to N'V$ and particle helicities $\oneh\to\lambda'_N,\lambda_V$ in the prescription B.}
\label{table5} 
\begin{center}
\begin{tabular}{r r c l}
\hline\hline
$\lambda'_N$ &  $\lambda_V$ & & $V(N,NV$)\\
\hline 
$+\oneh$&$+1$& &$\frac{\sqrt{2}k_L}{\sqrt{y}}\left[\frac{g}{1-y}+\frac{f}{2}\right]$\\
\\
$-\oneh$&$+1$& &$g\frac{\sqrt{2}(M_B-yM_N)}{\sqrt{y}}+\frac{f}{2M_N}
\frac{\sqrt{2}[k^2_\perp-(M_N+M_B)(1-y)(yM_N-M_B)]}{\sqrt{y}(1-y)}$\\
\\
$+\oneh$&$0$& &$g\frac{k^2_\perp+(1-y)^2M_NM_B-yM_V^2}{M_V\sqrt{y}(1-y)}$\\
 & & & $+\frac{f}{2M_N}
\frac{(yM_N-M_B)[k^2_\perp+y^2M_N^2-y(M_N^2+M_V^2+M_B^2)+M_B^2]}
{2M_Vy\sqrt{y}}$\\
\\
$-\oneh$&$0$& &$g\frac{(M_B-M_N)}{M_V\sqrt{y}}+\frac{f}{2M_N}\frac{k_R(1+y)[k^2_\perp-y(M_B^2+M_N^2+M^2_V)+M^2_B+y^2M_N^2]}{2M_Vy\sqrt{y}(1-y)}$\\
\\
$+\oneh$&$-1$& &$-g\frac{\sqrt{2}yk_R}{\sqrt{y}(1-y)}+\frac{f}{2M_N}\frac{\sqrt{2}k_R M_B}{\sqrt{y}}$\\
\\
$-\oneh$&$-1$& &$-\frac{f}{2M_N}\frac{\sqrt{2}k^2_R}{\sqrt{y}(1-y)}$\\
\hline\hline
\end{tabular}
\end{center}
\end{table}

The vertex function for the transition $N\rightarrow BV$ with the baryon $B$ being one of the octet states
is given by:
\begin{eqnarray}
V^{\lambda_N}_{\lambda'_B,\lambda_V}&=&
\tilde \phi_V^*\cdot \tilde T\,F_{NBV}(y,k_\perp)
\nonumber\\
& &\times\left[
g\,\bar u(p'_B,\lambda'_B)\gamma^\mu\,u(p_N,\lambda_N)
\varepsilon^*_{\mu,\lambda_V}-\frac{f}{2M_N}\bar u(p'_B,\lambda'_B)
i\sigma^{\mu\nu}p_{\nu,V}
\varepsilon^{*}_{\mu,\lambda_V}\right],\nonumber\\
& &\label{eq:vertex}
\end{eqnarray}
where $F_{NBV}$ is the vertex form factor, and the isospin factor is 
defined as 
\begin{eqnarray}
\bra{B}\tilde \phi_V^*\cdot \tilde T\ket{N}=(-1)^{\tau_V}
\frac{\bra{T_B}|\hat T|\ket{T_N}}{\sqrt{2T_B+2}}
\bra{T_N\tau_N1-\tau_V}T_B\tau_B\rangle.
\label{eq:isospin}
\end{eqnarray}
with $T_B=\oneh$, and $\bra{\oneh}|\hat T|\ket{\oneh}=\sqrt{6}$.

\begin{table}{}
\caption{\small Vertex functions for $N\to \Delta V$ and particle helicities $\oneh\to\lambda',\lambda_V$ in the prescription A.}
\label{table6} 
\begin{center}
\begin{tabular}{r r c l}
\hline\hline
$\lambda'$ &  $\lambda_V$ & & $V(N,NV$)\\
\hline 
$+\frac{3}{2}$&$+1$& &$-\frac{f}{M_V} \frac{k_L^2}{y\sqrt{y}(1-y)}$\\
\\
$+\frac{3}{2}$&$0$& &$\frac{f}{M_V}\frac{M_V\sqrt{2}k_L}{\sqrt{y}(1-y)}$\\
\\
$+\frac{3}{2}$&$-1$& &$\frac{f}{M_V}\frac{M_NM_B(1-y)^2-yM_V^2}{\sqrt{y}(1-y)}$\\
\\
$+\frac{1}{2}$&$1$& &$\frac{1}{\sqrt{3}}\frac{f}{M_V}\frac{k_L(k^2_\perp-2(1-y)M_B^2)}{M_By\sqrt{y}(1-y)}$\\
\\
$+\frac{1}{2}$&$0$& &$-\sqrt{\frac{2}{3}}\frac{f}{M_V}\frac{M_V[k^2_\perp-(1-y)M_B(M_B-yM_N)]}{M_B\sqrt{y}(1-y)}$\\
\\
$+\frac{1}{2}$&$-1$& &$-\frac{1}{\sqrt{3}}\frac{f}{M_V}\frac{k_R(-2M_NM_B(1-y)+yM^2_V)}{M_B\sqrt{y}(1-y)}$\\
\\
$-\frac{1}{2}$&$1$& &$\frac{1}{\sqrt{3}}\frac{f}{M_V}
\frac{2M_B(1-y)k^2_\perp-M^3_B(1-y)^2+M_NM^2_Vy^3}{M_By\sqrt{y}(1-y)}$\\
\\
$-\frac{1}{2}$&$0$& &$\sqrt{\frac{2}{3}}\frac{f}{M_V}\frac{M_Vk_R[M_Ny-(1-y)M_B]}{M_B\sqrt{y}(1-y)}$\\
\\
$-\frac{1}{2}$&$-1$& &$\frac{1}{\sqrt{3}}\frac{f}{M_V}\frac{M_Nk_R^2}{M_B\sqrt{y}(1-y)}$\\
\\
$-\frac{3}{2}$&$1$& &$\frac{f}{M_V}\frac{M_Bk_R(1-y)}{y\sqrt{y}}$\\
\\
$-\frac{3}{2}$&$0$& &$0$\\
\\
$-\frac{3}{2}$&$-1$& &$0$\\
\hline\hline
\end{tabular}
\end{center}
\end{table}

\begin{table}{}
\caption{\small Vertex functions for $N\to \Delta V$ and particle helicities $\oneh\to\lambda',\lambda_V$ in the prescription B.}
\label{table7} 
\begin{center}
\begin{tabular}{r r c l}
\hline\hline
$\lambda'$ &  $\lambda_V$ & & $V(N,\Delta V$)\\
\hline 
$+\frac{3}{2}$&$+1$& &$-\frac{f}{M_V}\frac{k_L^2}{y\sqrt{y}(1-y)}$\\
\\
$+\frac{3}{2}$&$0$& &$-\frac{f}{M_V}
\frac{k_L[k^2_\perp-yM^2_V-y(1-y)M_N^2+(1-y)M^2_B]}{\sqrt{2}M_Vy\sqrt{y}(1-y)}$\\
\\
$+\frac{3}{2}$&$-1$& &$\frac{f}{M_V}\frac{k^2_\perp+(1-y)(M_N+M_B)(M_B-yM_N)}{\sqrt{y}(1-y)}$\\
\\
$+\frac{1}{2}$&$1$& &$\frac{1}{\sqrt{3}}\frac{f}{M_V}\frac{k_L[k^2_\perp-2(1-y)M_B^2]}{M_By\sqrt{y}(1-y)}$\\
\\
$+\frac{1}{2}$&$0$& &$\frac{1}{\sqrt{6}}\frac{f}{M_V}\frac{[k^2_\perp+(1-y)M_B^2-y(1-y)M_N^2-yM^2_V)]
[k^2_\perp-(1-y)M_B^2+y(1-y)M_NM_B]}{M_BM_Vy\sqrt{y}(1-y)}$\\
\\
$+\frac{1}{2}$&$-1$& &$\frac{1}{\sqrt{3}}\frac{f}{M_V}
\frac{k_R[-k^2_\perp+(1-y)(yM_N^2-2M_NM_B-M_B^2)]}{M_B\sqrt{y}(1-y)}$\\
\\
$-\frac{1}{2}$&$1$& &$\frac{1}{\sqrt{3}}\frac{f}{M_V}\frac{
k^2_\perp[2M_B(1-y)-y^2M_N]+(1-y)[y^3M_N^3-M_NM^2_By^2-M_B^3(1-y)]}{M_By\sqrt{y}(1-y)}$\\
\\
$-\frac{1}{2}$&$0$& &$\frac{1}{\sqrt{6}}\frac{f}{M_V}
\frac{k_R[M_Ny-(1-y)M_B][y(1-y)M_N^2-(1-y)M^2_B+yM_V^2-k^2_\perp]}{M_BM_Vy\sqrt{y}(1-y)}$\\
\\
$-\frac{1}{2}$&$-1$& &$\frac{1}{\sqrt{3}}\frac{f}{M_V}\frac{M_Nk_R^2}{M_B\sqrt{y}(1-y)}$\\
\\
$-\frac{3}{2}$&$1$& &$\frac{f}{M_V} \frac{M_Bk_R(1-y)}{y\sqrt{y}}$\\
\\
$-\frac{3}{2}$&$0$& &$0$\\
\\
$-\frac{3}{2}$&$-1$& &$0$\\
\hline\hline
\end{tabular}
\end{center}
\end{table}

In the vertex function of Eq.~(\ref{eq:vertex}) there is an off-shell dependence introduced by the derivative coupling, leading to a freedom in defining the vertex momentum. One can consider the following two possibilities~\cite{HSS96}
\begin{eqnarray}
&{\mbox (A)}&\qquad p^{\nu}_{V}=(E_V,\vec p_V),\qquad \mbox{with}\quad E_V=\sqrt{m^2_V+\vec p_V^{\, 2}},\nonumber\\
&{\mbox (B)}&\qquad p^{\nu}_{V}=p_N-p'_B=(E_N-E'_B,\vec p_V).
\end{eqnarray}
In the following we will list the results for the vertex functions corresponding to both prescriptions,
although the calculation of the nucleon form factors is performed using the off-shell condition (B), as suggested in Ref.~\cite{HSS96}.

The results of the calculation for the particle helicities $\oneh\rightarrow\lambda'_N,\lambda_V$ are given in Table~\ref{table4} for prescription A and in Table~\ref{table5} for prescription B.

The corresponding results for helicity down of the nucleon are given by
\begin{eqnarray}
V_{\lambda',\lambda_V}^{-\oneh\,(N,BV)}(y,{\mathbf k}_\perp) 
&=&(-1)^{1/2+\lambda'+\lambda_V}\,
 V_{-\lambda',-\lambda_V}^{\oneh\,(N,BV)}(y,\hat{{\mathbf k}}_\perp),
 \nonumber
\end{eqnarray}
where $\hat{{\mathbf k}}_\perp = (k_x,-k_y)$.

The vertex function for the transition $N\rightarrow B V$ with the baryon being one of the decuplet states
is given by:
\begin{eqnarray}
V^{\lambda}_{\lambda',\lambda_V}&=&
\tilde \phi_V^*\cdot \tilde T\,F_{NBV}(y,k_\perp)
\nonumber\\
& &\times\frac{f}{M_V}
\,\bar u_\nu(\tilde p'_B,\lambda')\gamma_5\gamma_\mu\,u(\tilde p_N,
\lambda)
[p_V^\mu
\varepsilon^{\nu\,*}_{\lambda_V}-
p_V^\nu
\varepsilon^{\mu\,*}_{\lambda_V}]\nonumber\\
& &
\end{eqnarray}
where the isospin factor is 
defined as in Eq.~(\ref{eq:isospin}), with $T_B=\frac{3}{2}$ and 
$\bra{\oneh}|\hat T|\ket{\oneh}=2$.

The explicit results for particle helicities $\oneh\rightarrow\lambda',\lambda_V$ are given in Table~\ref{table6} for prescription A and in Table~\ref{table7} for prescription B.

The corresponding results for helicity down of the nucleon are given by
\begin{eqnarray}
V_{\lambda',\lambda_V}^{-\oneh\,(N,BV)}(y,{\mathbf k}_\perp) 
&=&(-1)^{3/2+\lambda'+\lambda_V}\,
 V_{-\lambda',-\lambda_V}^{\oneh\,(N,BV)}(y,\hat{{\mathbf k}}_\perp).
 \nonumber
\end{eqnarray}



\end{document}